\documentclass[a4paper,fleqn,usenatbib]{mnras}

\usepackage[T1]{fontenc}
\usepackage{ae,aecompl}

\usepackage{graphicx}	
\usepackage{amsmath}	
\usepackage{amssymb}	
\usepackage{subfigure}

\newcommand{\dinamo}{{\sc dinamo} }
\newcommand{\pcc}{\,{\rm cm}^{-3}}
\newcommand{\pcs}{\,{\rm cm}^{-2}}
\newcommand{\um}{\, {\rm \mu m}}

\newcommand{\kel}{\, {\rm K}}
\newcommand{\lsun}{\, {\rm L}_\odot}
\newcommand{\msun}{\, {\rm M}_\odot}

\newcommand{\nel}{n_{\rm e}}

\newcommand{\pc}{\, {\rm pc}}
\newcommand{\amin}{a_{\rm min}}
\newcommand{\amax}{a_{\rm max}}

\newcommand{\tel}{T_{\rm e}}

\newcommand{\ev}{\, {\rm eV}}
\newcommand{\kpc}{\, {\rm kpc}}
\newcommand{\geleven}{G$11.2$-$0.3$}
\newcommand{\gtwentyone}{G$21.5$-$0.9$}
\newcommand{\gtwentynine}{G$29.7$-$0.3$}
\newcommand{\gfiftyfour}{G$54.1$+$0.3$}
\newcommand{\yr}{\, {\rm yr}}

\title[Dust sizes in PWN]{Dust masses and grain size distributions of a sample of Galactic pulsar wind nebulae}

\author[Priestley et al.]{
F. D. Priestley$^{1}$, 
M. J. Barlow$^{2}$, 
I. De Looze$^{2,3}$
and H. Chawner$^{1}$
\\
$^{1}$School of Physics and Astronomy, Cardiff University, Queen's Buildings, The Parade, Cardiff CF24 3AA, UK \\
$^{2}$Department of Physics and Astronomy, University College London, Gower Street, London WC1E 6BT, UK\\
$^{3}$Sterrenkundig Observatorium, Ghent University, Krijgslaan 281 - S9, 9000 Gent, Belgium\\
}

\date{Accepted XXX. Received YYY; in original form ZZZ}

\pubyear{2019}

\begin{document}
\label{firstpage}
\pagerange{\pageref{firstpage}--\pageref{lastpage}}
\maketitle

\begin{abstract}

  We calculate dust spectral energy distributions (SEDs) for a range of grain sizes and compositions, using { physical properties} appropriate for five pulsar wind nebulae (PWNe) from which dust emission associated with the ejecta has been detected. By fitting the observed dust SED with our models, with the number of grains of different sizes as the free parameters, we are able to determine the grain size distribution and total dust mass in each PWN. We find that all five PWNe require large ($\ge 0.1 \um$) grains to make up the majority of the dust mass, with strong evidence for the presence of micron-sized or larger grains. Only two PWNe contain non-negligible quantities of small ($<0.01 \um$) grains. The size distributions are generally well-represented by broken power laws, although our uncertainties are too large to rule out alternative shapes. We find a total dust mass of $0.02-0.28 \msun$ for the Crab Nebula, depending on the composition and distance from the synchrotron source, in agreement with recent estimates. { For three objects in our sample, the PWN synchrotron luminosity is insufficient to power the observed dust emission, and additional collisional heating is required, either from warm, dense gas as found in the Crab Nebula, or higher temperature shocked material. For \gfiftyfour, the dust is heated by nearby OB stars rather than the PWN. Inferred dust masses vary significantly depending on the details of the assumed heating mechanism, but in all cases large mass fractions of micron-sized grains are required.}

\end{abstract}




\section{Introduction}

Core-collapse supernovae (CCSNe) have been proposed as a possible source for the large dust masses detected in high-redshift galaxies \citep{dunne2003,gall2011,gall2018}, due to the short lifetimes of their progenitors and their production of the elements which make up cosmic dust grains. Galaxy evolution models \citep{morgan2003,dwek2007,michalowski2010} require that a dust mass of $\sim 0.1 - 1.0 \msun$ per CCSNe must be injected into the interstellar medium (ISM) in order for SNe to account for the observed dust masses in the early Universe ($\sim 10^8 \msun$; \citealt{bertoldi2003}). Theoretical predictions for dust production by CCSNe (e.g. \citealt{todini2001,nozawa2003}) suggest that most, if not all, SNe should reach this range of values, and recent observations of both nearby supernova remnants (SNRs; \citealt{gomez2012,matsuura2015,delooze2017}) and extragalactic SNe \citep{bevan2017} appear to confirm this. However, the relevant quantity for the overall dust budget is not the total dust mass produced, but the amount which survives the passage of a reverse shock into the remnant.

Models of dust destruction in SNRs predict a wide range of survival rates, from complete destruction to almost total survival, depending on assumptions about the thermal and dynamical evolution of the gas, the grain size distribution and whether the dust is located in clumps \citep{bianchi2007,silvia2010,biscaro2016,bocchio2016,kirchschlager2019}. \citet{nozawa2007} found that grains with sizes $\lesssim 0.05 \um$ are destroyed completely by sputtering, whereas larger grains with $a \gtrsim 0.2 \um$ survive into the ISM. Similar conclusions about the importance of grain size to the destruction rate have been reached by other authors (e.g. \citealt{biscaro2016,micelotta2016}), and much of the variation in survival fractions between studies can be attributed to different assumptions about the grain size distributions.

Predicted size distributions of the dust formed in SNe ejecta also vary. \citet{nozawa2003} predicted that the size distributions of individual species would be approximately log-normal, while the overall size distribution could be fit by a broken power law, with the majority of the mass concentrated in the largest grains with sizes $\ge 0.1 \um$. \citet{bianchi2007} and \citet{bocchio2016} also found log-normal distributions, but with most of the dust mass in smaller grains. The predicted size distributions can vary significantly depending on the level of clumping assumed, as well as on the specific ejecta properties (e.g. \citealt{sarangi2015}), with large hydrogen envelopes at the time of explosion tending to result in larger grain sizes \citep{kozasa2009} in dust formation models.

Observationally, constraints on grain sizes are limited. \citet{gall2014} inferred that large, micron-sized or greater grains were needed to reproduce the wavelength-dependent extinction in SN 2010jl. \citet{wesson2015} and \citet{bevan2016} both found evidence for large ($>0.1 \um$) grain sizes in SN 1987A at late times, while \citet{bevan2017} required similarly large grain sizes in SN 1980K and SN 1993J, but not Cassiopeia A (Cas A). \citet{temim2013} and \citet{owen2015} used different physical models to fit the observed dust spectral energy distribution (SED) of the Crab Nebula with power law grain size distributions, agreeing on a required maximum grain size above $0.1 \um$ but { disagreeing on the power law index, with \citet{temim2013} finding power law indices larger than the MRN value of $3.5$ \citep{mathis1977}, and \citet{owen2015} generally finding shallower size distributions. \citet{owen2015} found a minimum grain size of order $\amin \sim 0.01 \um$ for most geometries and dust compositions, while \citet{temim2013} were not able to constrain this parameter in their models.} \citet{priestley2019} assumed an MRN distribution in modelling the Cassiopeia A dust SED, but found the results were not particularly sensitive to different choices of power law parameters.

All of the previous studies assume either a single size of dust grain, or that the grain size distribution is a power law, despite theoretical predictions that the size distribution of newly-formed dust should be log-normal. In this paper, we attempt to determine the grain size distribution for a sample of Galactic pulsar wind nebulae (PWN) - SNRs with strong synchrotron emission powered by a central pulsar - with confirmed ejecta dust emission, without assuming any particular functional form. { Compared to strongly interacting SNRs such as Cas A, PWNe are much more homogenous, and the properties responsible for dust heating (radiation field, gas density and temperature) can be reasonably treated as constant.} As the grain temperature as a function of radius is then, in principle, well determined, the SED can be modelled with the mass of dust grains of different sizes as the only fitting parameters, providing a measurement of both the total dust mass and the grain size distribution.

\section{Method}

We calculate the emitted SED from single dust grains of different sizes using \dinamo \citep{priestley2019}, which takes as input the dust optical and physical properties and the local gas density, temperature and radiation field, and returns the equilibrium temperature distribution for each grain and the resulting grain emissivity. We consider three grain species - MgSiO$_3$, with optical constants from \citet{dorschner1995} and \citet{laor1993}, and two varieties of amorphous carbon, ACAR and BE, from \citet{zubko1996}, with the optical constants extended to shorter wavelengths using data from \citet{uspenskii2006}. The method of combining the optical constants is described in \citet{priestley2019} for MgSiO$_3$ and \citet{owen2015} for the carbon species. The dust properties used are given in Table \ref{tab:dustprop}. We obtain single-grain SEDs for grains of radius $0.001$, $0.01$, $0.1$ and $1.0 \um$, for each dust species and PWN.

\begin{table}
  \centering
  \caption{Dust species and their adopted densities $\rho_g$, sublimation temperatures $T_{\rm sub}$ and references for the optical constants. References: (1) \citet{dorschner1995} (2) \citet{laor1993} (3) \citet{zubko1996} (4) \citet{uspenskii2006}.}
  \begin{tabular}{cccc}
    \hline
    Dust species & $\rho_g$/${\rm g} \pcc$ & $T_{\rm sub}$/$\kel$ & $n$-$k$ \\
    \hline
    MgSiO$_3$ & $2.5$ & $1500$ & (1),(2) \\
    Am. carbon ACAR & $1.6$ & $2500$ & (3),(4) \\
    Am. carbon BE &  $1.6$ & $2500$ & (3),(4) \\
    \hline
  \end{tabular}
  \label{tab:dustprop}
\end{table}

The Crab Nebula is the closest and best-studied object in our sample of PWN, and as such has much more reliably determined parameters for our modelling technique. The commonly-used value for the distance to the Crab is $2 \kpc$ \citep{trimble1968}. However, a Gaia measurement of the pulsar's parallax \citep{fraser2019} gives a larger distance of $3.37^{+4.04}_{-0.97} \kpc$. We use the \citet{trimble1968} distance for consistency with previous work - using the larger distance would increase all dust masses by a factor of $\sim 2.8$. The luminosity and SED of the PWN synchrotron emission are given by \citet{hester2008}. The SNR radius is $2 \pc$ \citep{hester2008}, with X-ray emission concentrated in the inner $0.5 \pc$. Previous studies of the dust emission have used radii for the dust location ranging from $0.5 \pc$ \citep{temim2013} to $2.0 \pc$ \citep{owen2015}, so we consider distances within this range and determine the radiation field treating the PWN as a point source. For the collisional heating we assume the dust is located within the dense knots, with densities and temperatures of $\nel = 10^4 \pcc$ and $\tel = 3000 \kel$ inferred from molecular observations \citep{loh2012,richardson2013,priestley2017}. The most common element in the Crab ejecta is helium \citep{macalpine2008,owen2015}, so we include collisional heating by helium nuclei with the same density and temperature as the electrons, although this is unlikely to be a significant heating mechanism. These properties are listed in Table \ref{tab:snrprop}. Finally, we use observed SNR dust fluxes from \citet{delooze2019}, based on a multi-component fit accounting for ISM dust, synchrotron emission and an unidentified source of excess millimetre emission, combined with the SCUBA upper limit from \citet{gomez2012}. These are listed in Table \ref{tab:dustflux}.

\begin{table*}
  \centering
  \caption{{ Distance, radius, radiation field SED and luminosity, assumed distance $d$ of dust from the heating source and electron temperature and density for our sample of PWN. (* value adopted from Crab Nebula)}}
  \begin{tabular}{cccccccc}
    \hline
    Parameter & Crab Nebula & \geleven & \gtwentyone & \gtwentynine & \multicolumn{3}{c}{\gfiftyfour} \\
    \hline
    Distance / $\kpc$ & $2.0$ & $4.4$ & $4.7$ & $5.8$ & \multicolumn{3}{c}{$6.2$} \\
    Radius / $\pc$ & $2.0$ & $2.4$ & $2.3$ & $2.0$ & \multicolumn{3}{c}{$2.3$} \\
    SED & \citet{hester2008} & Power law & Power law & Power law & Power law & $30 \, 000 \kel$ BB $2 \pc$ & $30 \, 000 \kel$ BB $0.2 \pc$ \\
    $L_{\rm tot}$ / erg s$^{-1}$ & $1.3 \times 10^{38}$ & $2.4 \times 10^{35}$ & $1.6 \times 10^{36}$ & $1.5 \times 10^{36}$ & $3.7 \times 10^{35}$ & $1.1 \times 10^{38}$ & $1.1 \times 10^{38}$ \\
    $d$ / $\pc$ & $0.5-2$ & $2$ & $2$ & $2$ & $2$ & $2$ & $0.2$ \\
    *$\tel$ / $\kel$ & $3000$ & $3000$ & $3000$ & $3000$ & $3000$ & $3000$ & $3000$ \\
    *$\nel$ / $\pcc$ & $10^4$ & $10^4$ & $10^4$ & $10^4$ & $10^4$ & $10^4$ & $10^4$ \\
    \hline
  \end{tabular}
  \label{tab:snrprop}
\end{table*}

\begin{table*}
  \centering
  \caption{SNR dust fluxes for our sample of SNRs. References: \citet{gomez2012}, \citet{delooze2019} - Crab Nebula; \citet{chawner2019} - \geleven, \gtwentyone; \citet{chawner2019}, \citet{temim2019} - \gtwentynine; \citet{temim2017} - \gfiftyfour}
  \begin{tabular}{cccccc}
    \hline
     & \multicolumn{5}{c}{$F_\nu / {\rm Jy}$} \\
    Waveband & Crab Nebula & \geleven & \gtwentyone & \gtwentynine & \gfiftyfour \\
    \hline
    IRAC $8 \um$ & $0.05 \pm 0.13$ & - & - & - & - \\
    WISE $22 \um$ & $17.8 \pm 3.7$ & - & - & - & - \\
    MIPS $24 \um$ & $20.9 \pm 2.9$ & $5.6 \pm 0.3$ & $0.0 \pm 0.02$ & $0.19 \pm 0.03$ & $28.6 \pm 2.6$ \\
    PACS $70 \um$ & $168.2 \pm 19.5$ & $47.7 \pm 6.7$ & $3.7 \pm 0.3$ & $4.4 \pm 1.5$ & $87.9 \pm 11.4$ \\
    PACS $100 \um$ & $142.2 \pm 18.5$ & - & - & $5.1 \pm 1.7$ & $68.8 \pm 13.4$ \\
    PACS $160 \um$ & $69.9 \pm 14.0$ & $71.9 \pm 15.7$ & $6.0 \pm 0.7$ & $1.6 \pm 1.9$ & $29.0 \pm 14.9$ \\
    SPIRE $250 \um$ & $25.1 \pm 6.7$ & $26.6 \pm 5.5$ & $2.2 \pm 0.9$ & $0.52 \pm 2.10$ & $6.9 \pm 5.2$ \\
    SPIRE $350 \um$ & $10.4 \pm 6.0$ & $10.1 \pm 3.0$ & $0.6 \pm 0.8$ & $0.16 \pm 1.15$ & $1.6 \pm 2.8$ \\
    SPIRE $500 \um$ & $3.7 \pm 6.0$ & $2.3 \pm 0.9$ & $0.0 \pm 0.4$ & $0.00 \pm 0.20$ & $0.4 \pm 1.1$ \\
    SCUBA $850 \um$ & $0.0 \pm 19.0$ & - & - & - & - \\
    \hline
  \end{tabular}
  \label{tab:dustflux}
\end{table*}

Of the other PWNe in our study, three (\geleven, \gtwentyone \, and \gtwentynine) were studied in-depth by \citet{chawner2019} after being identified as containing dust associated with the ejecta { (dust emission from \gtwentynine \, was originally reported by \citet{temim2017b})}. The background- and synchrotron-subtracted fluxes are listed in Table \ref{tab:dustflux}. { For \gtwentynine \, we use the PACS fluxes from \citet{temim2019} with line contributions subtracted (those authors use a slightly smaller aperture to determine the fluxes, but the differences in flux are much smaller than the uncertainties). No line corrections are available for the other two PWNe, so the true dust fluxes - particularly at $160 \um$ - may be lower than the listed values.} The adopted distances are $4.4 \kpc$ (\geleven ; \citealt{green2004}), $4.7 \kpc$ (\gtwentyone ; \citealt{camilo2006}) { and $5.8 \kpc$ (\gtwentynine ; \citealt{verbiest2012})}, while \citet{chawner2019} give the radii as $2.4 \pc$, $2.3 \pc$ and $3.7 \pc$ respectively { (for \gtwentynine, \citet{chawner2019} used a distance of $10.6 \kpc$ from \citet{su2009}, so for our adopted distance the radius is $2.0 \pc$)}.

\begin{figure}
  \centering
  \includegraphics[width=\columnwidth]{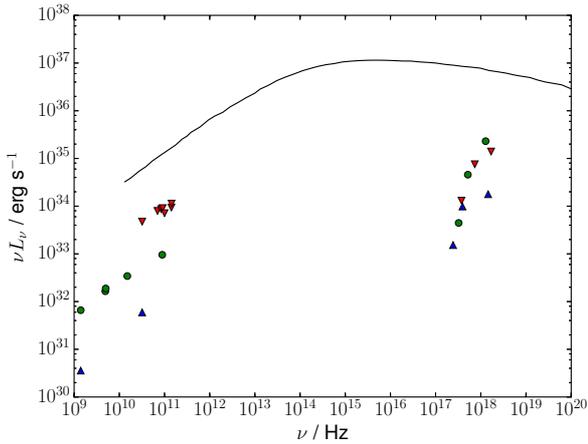}
  \caption{Crab Nebula SED from \citet{hester2008} compared with radio and X-ray luminosities for \geleven \, (blue upward triangles), \gtwentyone \, (red downward triangles) and \gtwentynine \, (green circles).}
  \label{fig:syncfit}
\end{figure}

\begin{figure}
  \centering
  \includegraphics[width=\columnwidth]{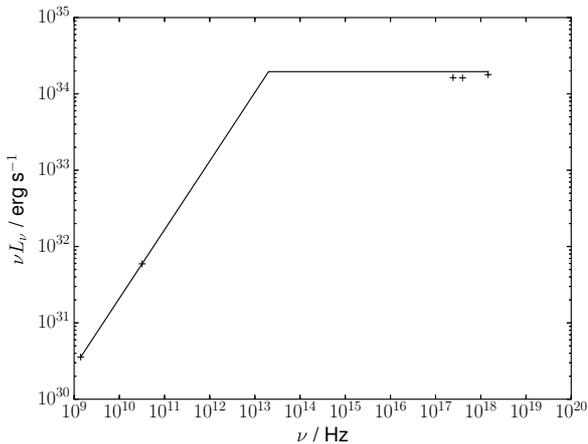}
  \caption{Observed radio and X-ray luminosities for \geleven \, (black crosses), assuming an absorbing column density of $N_{\rm H} = 5.1 \times 10^{22} \pcs$, and a two-component power law fit.}
  \label{fig:plfit}
\end{figure}

Unlike the Crab, detailed information on the synchrotron SED is unavailable - radio and X-ray data can be used to constrain the behaviour at the extremes, but not the optical-UV part, which is most important for dust heating. Figure \ref{fig:syncfit} shows the radio luminosities from \citet{chawner2019}, and X-ray data from {\it Chandra} observations { extracted from the same apertures}, for the three PWNe, and the Crab Nebula SED from \citet{hester2008} as a comparison. While the radio data are similar to the Crab SED, although less luminous, the X-ray luminosities increase with energy rather than the flat trend seen in the Crab. This can be attributed to absorption in the ISM, which affects lower-energy photons more strongly. As PWNe are generally found to approximately follow a $F_{\nu} \propto \nu^{-1}$ relation in the X-ray region \citep{gaensler2006}, we assume this is the underlying relation, and determine the necessary hydrogen column density towards the remnant $N_{\rm H}$ to recover this trend, using cross-sections from \citet{verner1995} and assuming a solar composition \citep{asplund2009}. We then fit the SED using a two-component power law, taking the radio indices from \citet{chawner2019} and an X-ray index of $-1$. The parameters are given in Table \ref{tab:plfits}, and the result for \geleven \, is shown in Figure \ref{fig:plfit}. We assume dust distances from the PWN source as listed in Table \ref{tab:snrprop}, and the same electron density and temperature as the Crab.

\begin{table*}
  \centering
  \caption{Parameters for PWN synchrotron SED power law fits. The SED is approximated as $L_{\nu} \propto \nu^{-\alpha_1}$ for $\nu \le \nu_0$ and $L_{\nu} \propto \nu^{-\alpha_2}$ for $\nu > \nu_0$.}
  \begin{tabular}{ccccc}
    \hline
    Parameter & \geleven & \gtwentyone & \gtwentynine & \gfiftyfour \\
    \hline
    $N_{\rm H}$ / $10^{23} \pcs$ & $0.51$ & $1.9$ & $2.0$ & $1.3$ \\
    $\nu_0$ / $10^{13}$ Hz & $2.0$ & $5.6$ & $140$ & $0.7$ \\
    $\nu_0 L_{\nu}(\nu_0)$ / $10^{34}$ erg s$^{-1}$ & $1.9$ & $12.7$ & $17.3$ & $2.8$ \\
    $\alpha_1$ & $0.10$ & $0.56$ & $0.43$ & $0.16$ \\
    $\alpha_2$ & $1.0$ & $1.0$ & $1.0$ & $1.0$ \\
    \hline
  \end{tabular}
  \label{tab:plfits}
\end{table*}

{ Our $N_{\rm H}$ values used to determine the true X-ray luminosities are higher than those obtained by other authors using more sophisticated models, ranging from factors of $\sim 2$ for \geleven \, \citep{borkowski2016} to over an order of magnitude for the other two PWNe \citep{temim2012,guest2019}. Additionally, PWN synchrotron emission is not necessarily well described by a double power law, particularly at higher energies where there may be multiple spectral breaks. However, the dust SEDs produced by this method are indistinguishable from those generated using theoretical PWN spectra for the individual remnants \citep{tanaka2011,gelfand2014}, so we consider the approximation acceptable.}

The final PWN we consider, \gfiftyfour, is located at a distance of $6.2 \kpc$ \citep{leahy2008}, which, for an angular size of $1.3'$ gives an SNR radius of $2.3 \pc$. We use {\it Spitzer} and {\it Herschel} fluxes taken from \citet{temim2017}, as listed in Table \ref{tab:dustflux}. Unlike the previous four objects, it has been suggested that the dust observed in \gfiftyfour \, is heated by nearby OB stars, rather than the PWN synchrotron emission (\citealt{temim2017}, although see \citealt{rho2018} for a counterargument). Using the same method as for the previous three PWN, we determine a two-power law fit to the radio \citep{rho2018} and X-ray data, and we model \gfiftyfour \, assuming this is the source of the dust heating. However, we also consider heating by a blackbody with a temperature of $30 \, 000 \kel$ and a luminosity of $30 \, 000 \lsun$ to represent a typical O star. { We assume the synchrotron heating source is located $2 \pc$ from the dust. For heating by OB stars, the situation is more complicated, as there are multiple stars in the vicinity of the PWN so a single distance is inappropriate. We investigated distances of $0.2 \pc$ and $2 \pc$ to cover a plausible range of values and constrain the possible dust properties - determining these more accurately would require detailed three-dimensional modelling beyond the scope of this paper.}

For each PWN, we obtain single-grain SEDs for grains of radius $0.001$, $0.01$, $0.1$ and $1.0 \um$, for each dust species. We then convolve the SED with the appropriate filter profiles and fit the observed fluxes - listed in Table \ref{tab:dustflux} - using the Monte Carlo Markov Chain (MCMC) code {\it emcee} \citep{foreman2013}, with the number of grains of each radius as the four free parameters, and flat priors in log space over the range $\log N(a) = 40-52$ for each grain size, which covers the range in which dust grains can contribute significantly to the SED while not exceeding constraints on the maximum flux. We use $300$ walkers with $5000$ steps per walker, burning the first $100$ steps, which we find converges for all models. This method results in the probability density as a function of $\log N(a)$ for each grain size, which generally shows a clear peak unless emission from the grain size in question is insignificant. However, the formally best-fitting values of $N(a)$ can differ significantly from these peaks, as particular values can combine to give a fractionally better (if physically meaningless) $chi^2$ value. As such, we present the average values of $\log N(a)$ over all MCMC runs, which we find to be a more accurate indicator of the probability distribution.

\section{Results}

\subsection{The Crab Nebula}

The Crab Nebula dust masses returned by our models for each grain size are listed in Table \ref{tab:crabdust}, with error bars corresponding to the 16th and 84th percentile range. Formally, our best-fit model is for ACAR grains at $0.5 \pc$. However, with the exception of ACAR $2.0 \pc$ and BE and MgSiO$_3$ $0.5 \pc$, all models fit the data well, and we are not able to meaningfully distinguish between them. We find total dust masses ranging from $0.02-0.28 \msun$, depending on grain species and the distance from the PWN. For carbon grains, our results are similar to those of \citet{delooze2019}, but our masses are an order of magnitude higher for MgSiO$_3$. \citet{delooze2019} found best-fit cold dust temperatures of $35-50 \kel$, whereas our $1.0 \um$ grain temperatures are $\sim 25 \kel$, with the lower temperatures being sufficient to explain our higher masses. Our dust masses are a factor of $3-4$ less than those found by \citet{gomez2012} and \citet{owen2015}, as we adopted smaller SNR far-IR fluxes derived by \citet{delooze2019}. They are mostly consistent with those of \citet{temim2013}, { who used the higher far-IR fluxes from \citet{gomez2012} but whose best-fit models predicted values closer to the \citet{delooze2019} values.} \citet{nehme2019} also found that the far-IR dust fluxes from the Crab are lower than those used by \citet{gomez2012}, using an independent method from \citet{delooze2019}, and determined a dust mass of $0.06 \pm 0.04 \msun$, consistent with our results.

\begin{table*}
  \centering
  \caption{Mass of dust grains of different sizes, total dust masses and $\chi^2$ values for different grain species and distances from the PWN radiation source in the Crab Nebula.}
  \begin{tabular}{c|cccc|cc}
    \hline
    & \multicolumn{4}{|c|}{$\log M_{\rm dust}(a)$/$\msun$} &  \\
    Model & $0.001 \um$ & $0.01 \um$ & $0.1 \um$ & $1.0 \um$ & $M_{\rm tot}$/$\msun$ & $\chi^2$ \\
    \hline
    ACAR $0.5 \pc$ & $-5.85^{+0.06}_{-6.08}$ & $-3.28^{+0.23}_{-3.96}$ & $-2.63^{+0.41}_{-3.30}$ & $-1.64^{+0.11}_{-0.21}$ & $0.026^{+0.004}_{-0.006}$ & $1.11$ \\
    ACAR $1.0 \pc$ & $-4.46^{+0.58}_{-7.18}$ & $-2.42^{+0.15}_{-5.01}$ & $-2.15^{+0.49}_{-3.58}$ & $-1.55^{+0.20}_{-1.13}$ & $0.039^{+0.010}_{-0.014}$ & $2.03$ \\
    ACAR $2.0 \pc$ & $-3.94^{+0.34}_{-6.81}$ & $-1.96^{+0.14}_{-4.36}$ & $-2.07^{+0.60}_{-4.26}$ & $-1.71^{+0.23}_{-1.61}$ & $0.039^{+0.012}_{-0.021}$ & $4.20$ \\
    BE $0.5 \pc$ & $-5.80^{+0.07}_{-6.13}$ & $-3.29^{+0.23}_{-4.24}$ & $-2.78^{+0.42}_{-3.27}$ & $-1.53^{+0.06}_{-0.10}$ & $0.032^{+0.003}_{-0.004}$ & $3.99$ \\
    BE $1.0 \pc$ & $-4.52^{+0.26}_{-7.21}$ & $-2.42^{+0.14}_{-4.10}$ & $-2.20^{+0.52}_{-4.01}$ & $-1.34^{+0.15}_{-0.98}$ & $0.056^{+0.013}_{-0.026}$ & $1.64$ \\
    BE $2.0 \pc$ & $-3.97^{+0.59}_{-7.23}$ & $-2.00^{+0.15}_{-4.90}$ & $-1.93^{+0.60}_{-4.32}$ & $-1.27^{+0.17}_{-1.17}$ & $0.076^{+0.017}_{-0.022}$ & $2.59$ \\
    MgSiO$_3$ $0.5 \pc$ & $-6.27^{+0.07}_{-5.53}$ & $-3.24^{+0.18}_{-3.33}$ & $-2.54^{+0.50}_{-3.54}$ & $-1.14^{+0.06}_{-0.12}$ & $0.076^{+0.009}_{-0.012}$ & $4.25$ \\
    MgSiO$_3$ $1.0 \pc$ & $-5.29^{+0.15}_{-6.29}$ & $-2.39^{+0.13}_{-3.05}$ & $-1.84^{+0.51}_{-4.03}$ & $-0.90^{+0.16}_{-1.40}$ & $0.144^{+0.043}_{-0.083}$ & $1.95$ \\
    MgSiO$_3$ $2.0 \pc$ & $-4.11^{+0.59}_{-7.18}$ & $-1.86^{+0.13}_{-4.60}$ & $-1.63^{+0.55}_{-4.24}$ & $-0.74^{+0.16}_{-1.00}$ & $0.218^{+0.059}_{-0.099}$ & $2.15$ \\
    \hline
  \end{tabular}
  \label{tab:crabdust}
\end{table*}

All our models require a large fraction of the total dust mass to be in $1.0 \um$ grains, with the fraction in smaller grains increasing with distance from the PWN but remaining below $50 \%$ in all cases. Smaller grain sizes ($\le 0.01 \um$) do not contribute more than $\sim 0.01 \msun$ in any model to the total dust mass, and the smallest grain size has a mass consistent with zero in all models, contributing an insignificant amount to the total SED. The requirement for micron-sized grains is in agreement with \citet{temim2013} and \citet{owen2015} - however, \citet{temim2013} claimed to find no constraint on the smallest grain size in their models. We find that the mass of dust in the smallest grain size considered, $0.001 \um$, is strongly constrained for all models, as these grains emit efficiently in the mid-IR due to stochastic heating, which was not treated by \citet{temim2013}. The low observed fluxes at these wavelengths put a strict upper limit on the number of grains of this size that can be present.

\begin{figure}
  \centering
  \includegraphics[width=\columnwidth]{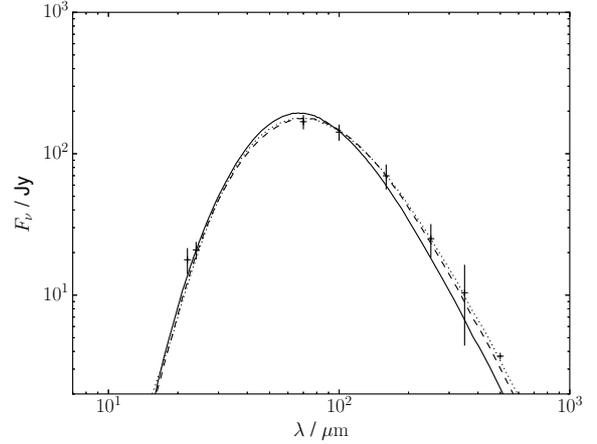}
  \caption{Model SEDs for $d=0.5 \pc$ (solid line), $1.0 \pc$ (dashed line) and $2.0 \pc$ (dotted line) for BE grains, with Crab Nebula SNR dust fluxes from \citet{delooze2019} (black crosses).}
  \label{fig:crabbeflux}
\end{figure}

\begin{figure}
  \centering
  \includegraphics[width=\columnwidth]{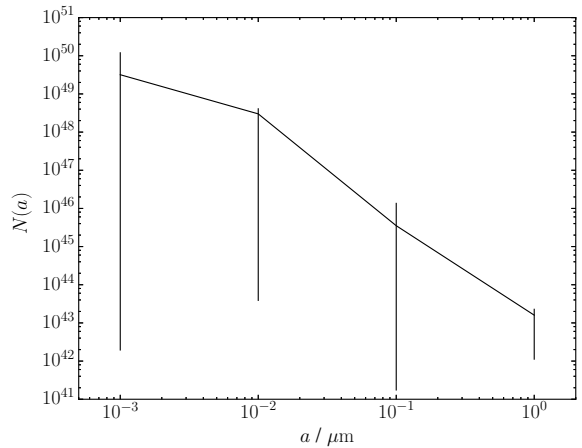}
  \caption{Number of grains versus grain size for BE grains at $d=2 \pc$ in the Crab Nebula.}
  \label{fig:crabbesize}
\end{figure}

Figure \ref{fig:crabbeflux} shows the model SEDs for BE grains at various distances, and Figure \ref{fig:crabbesize} shows the grain size distribution for the $2.0 \pc$ model. For this grain species, the distance from the heating source does not affect the ability of the model to fit the data - closer distances require a larger proportion of the mass to reside in micron-sized, as opposed to $0.1 \um$, grains, while larger distances require an increased number of $0.01 \um$ grains to provide the mid-IR flux. For the $0.5 \pc$ model, the number of $1.0 \um$ grains is tightly constrained, while for the other models significant variations in the mass in any one grain size are allowed. However, the majority of the mass is always contained in grains with radii $\ge 0.1 \um$. The average values resemble a power law with a reduced number of grains at small radii, such as that produced by fast non-radiative shocks due to sputtering \citep{dwek1996}.

\begin{figure*}
  \centering
  \subfigure{\includegraphics[width=\columnwidth]{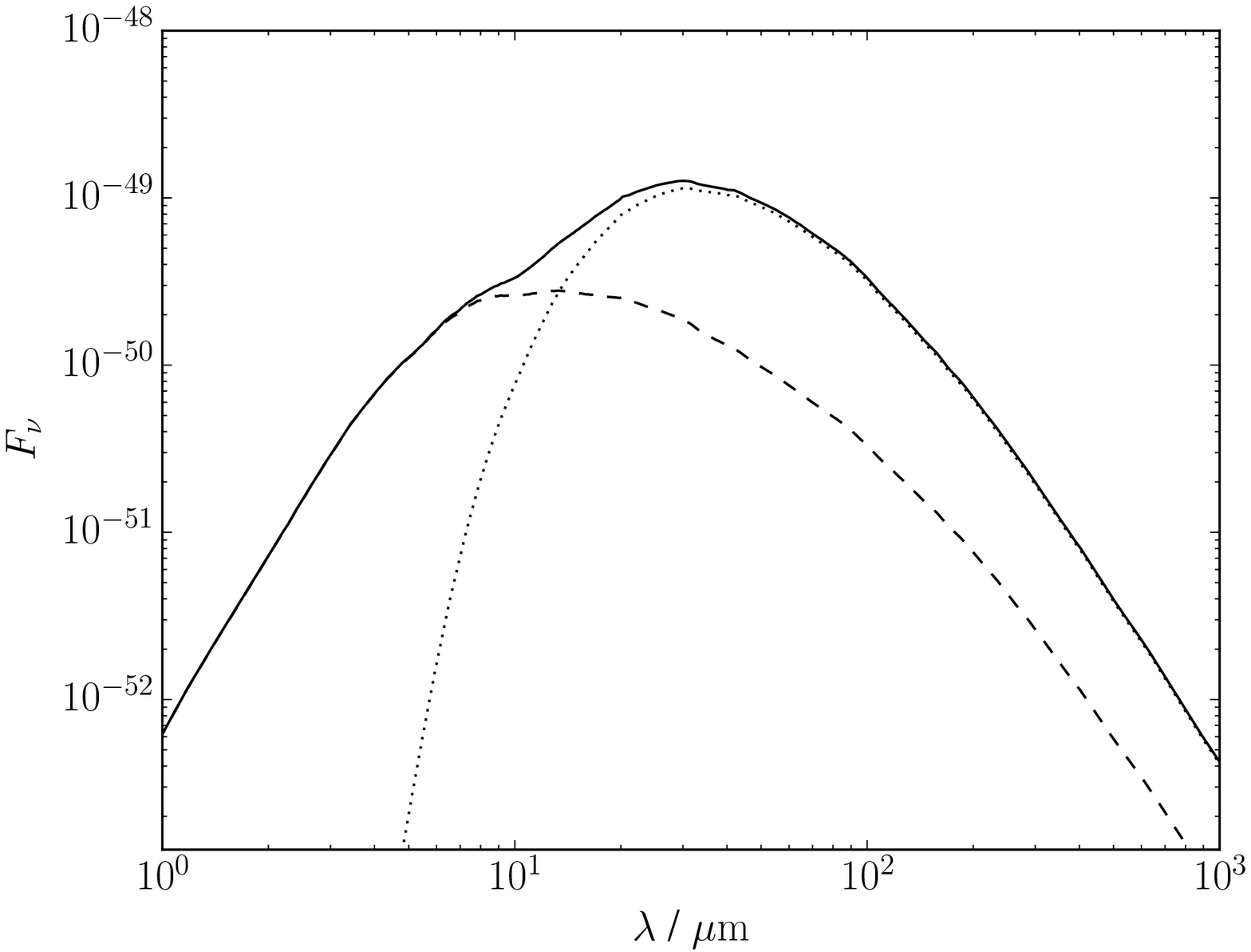}}\quad
  \subfigure{\includegraphics[width=\columnwidth]{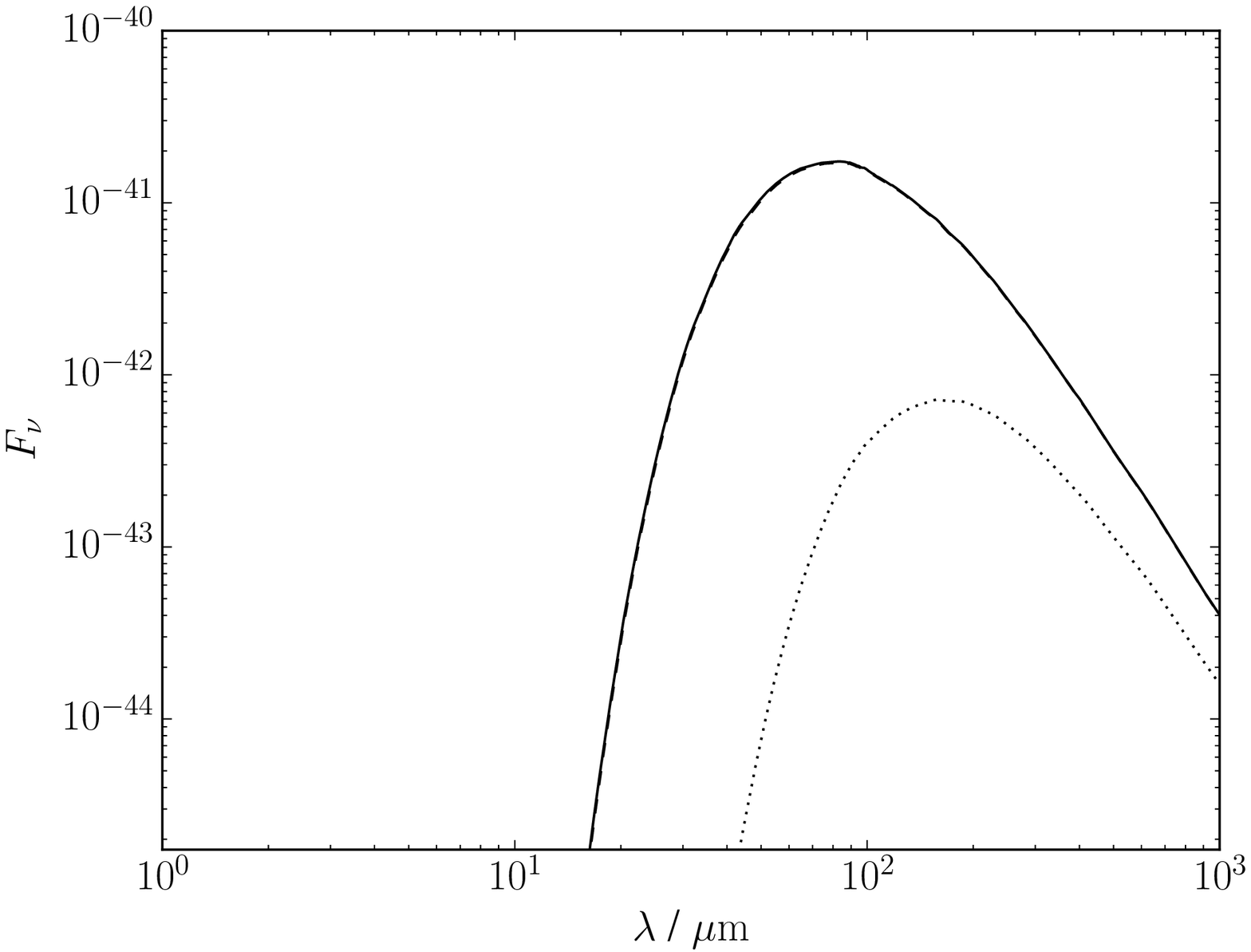}}
  \caption{Crab Nebula dust SEDs for ACAR grains at $0.5 \pc$ from the PWN heated by only the radiation field (dashed lines), only particle collisions (dotted lines) or both (solid lines), for a grain size of $0.001 \um$ (left panel) or $1.0 \um$ (right panel).}
  \label{fig:sizeheat}
\end{figure*}

Previous models of dust heating in the Crab Nebula \citep{temim2013,owen2015} have only taken into account radiative heating by the PWN flux, whereas our models also include heating by electrons and ions in the ambient medium. Figure \ref{fig:sizeheat} shows the emitted SEDs of ACAR grains at $0.5 \pc$ from the PWN, heated by one or both of the two heating mechanisms, for grains of radius $0.001$ and $1.0 \um$. For the larger grain size, the dust heating is virtually entirely radiative - the contribution by particle heating makes no difference to the emitted SED. The smaller grains, by contrast, show significantly different behaviour when collisional heating is included. These grains are stochastically heated by photons, causing the emission to be dominated by the transiently heated grains while the majority are at much cooler temperatures and do not contribute much. The addition of particle collisions raises the minimum temperature of the grains from $\sim 20$ to $40 \kel$, substantially increasing the emission at longer wavelengths. However, as the majority of the emission in our models comes from the largest grains, the overall effect of collisional heating on the dust emission is small - for the BE $1.0 \pc$ model, radiative heating accounts for $84 \%$ of the emitted energy, with electron collisions supplying the remaining $16 \%$.

For the environmental parameters, we vary the distance from the PWN radiation source but assume the dust is entirely located in the H$_2$ emitting clumps. It is also possible that some (or all) of the dust is in the hotter photoionized gas outside the clumps. Taking typical values for this phase as $\nel = 10 \pcc$ and $\tel = 10^4 \kel$, we rerun our BE $1.0 \pc$ model with these parameters - however, we find the change in the results to be negligible. This is unsurprising, as we noted that particle collisions supply a small fraction of the overall emitted energy, particularly for large grains.

Although we treat the number of grains of each size as a free parameter, the grain radii themselves are fixed. Our models show a clear preference for the largest grain sizes investigated, while the smallest sizes are essentially ruled out as a significant presence. We therefore investigate possible variations in the maximum grain radius, from $0.5$ to $10 \um$. For the BE $1.0 \pc$ model, the agreement with observations worsens as the maximum size is increased from $1.0 \um$, becoming noticeably discrepant for grain sizes beyond $2 \um$. The total dust mass increases to $0.080 \msun$ for $\amax = 2 \um$, and decreases to $0.041 \msun$ for $\amax = 0.5 \um$.

In addition to the thermal electrons, the Crab Nebula is subjected to a flux of charged particles from the PWN, leading to ionization rates $\sim 10^7$ times higher than typical ISM values \citep{richardson2013,priestley2017}, which could plausibly affect the dust temperature balance. \citet{richardson2013} estimate a maximum energy density of ionizing particles of $2000 \ev \pcc$, which is comparable to the thermal energy density ($k_b \tel \nel \sim 2500 \ev \pcc$ for our parameters). However, given the relatively unimportant role of collisional heating on our results, and the fact that the energy spectrum of electrons peaks in the MeV range \citep{atoyan1996}, where only a small fraction of the energy is deposited upon collisions with dust grains \citep{barlow1978,dwek1987}, we consider it justified to neglect this process.

\subsection{\geleven, \gtwentyone \, and \gtwentynine}

The dust masses for the three SNRs { with no external heating source}, \geleven, \gtwentyone \, and \gtwentynine, are listed in Table \ref{tab:pwndust}. { \geleven \, and \gtwentyone \, are found to contain significantly more dust than the Crab Nebula - $0.2-2.3$ and $0.04-0.32 \msun$ respectively, depending on the grain composition, whereas \gtwentynine \, contains a similar quantity ($0.005-0.14 \msun$).} However, the distribution of mass amongst grain sizes remains similar, with micron-sized grains containing the majority of the mass { in all cases}, and the fraction of mass in grains $> 0.1 \um$ essentially unity. $0.001 \um$ grains contribute negligibly to both the mass and the total dust emission except for in \geleven, where they are required to produce the observed $24 \um$ flux { (although this data point may be contaminated by line emission, or have a greater synchrotron contribution than assumed by our power law fit)}. While the dust masses vary quite significantly depending on the assumed grain composition, the SEDs and size distributions are not greatly affected. MgSiO$_3$ grains result in lower $\chi^2$ values for all three SNRs, although the differences are not great enough to conclusively favour silicate grains over carbon.

\begin{table*}
  \centering
  \caption{Mass of dust grains of different sizes, total dust masses and $\chi^2$ values for different grain species in \geleven, \gtwentyone \, ({ with and without the $160 \um$ flux)} and \gtwentynine.}
  \begin{tabular}{c|cccc|cc}
    \hline
    & \multicolumn{4}{|c|}{$\log M_{\rm dust}(a)$/$\msun$} &  \\
    Model & $0.001 \um$ & $0.01 \um$ & $0.1 \um$ & $1.0 \um$ & $M_{\rm tot}$/$\msun$ & $\chi^2$ \\
    \hline
    \geleven \, ACAR & $-3.14^{+0.15}_{-4.15}$ & $-1.96^{+0.43}_{-6.27}$ & $-0.85^{+0.24}_{-3.95}$ & $-0.73^{+0.39}_{-2.57}$ & $0.339^{+0.145}_{-0.101}$ & $8.14$ \\
    \geleven \, BE & $-3.22^{+0.22}_{-5.37}$ & $-1.91^{+0.40}_{-5.79}$ & $-0.99^{+0.33}_{-4.39}$ & $-0.25^{+0.16}_{-0.26}$ & $0.674^{+0.171}_{-0.156}$ & $7.74$ \\
    \geleven \, MgSiO$_3$ & $-3.06^{+0.20}_{-5.94}$ & $-1.64^{+0.42}_{-5.81}$ & $-0.99^{+0.39}_{-4.63}$ & $0.24^{+0.10}_{-0.11}$ & $1.861^{+0.397}_{-0.372}$ & $5.46$ \\
    \hline
    \gtwentyone \, ACAR & $-6.75^{+0.04}_{-5.29}$ & $-4.11^{+0.49}_{-5.20}$ & $-1.73^{+0.07}_{-0.02}$ & $-1.38^{+0.12}_{-0.37}$ & $0.061^{+0.014}_{-0.021}$ & $5.31$ \\
    \gtwentyone \, BE & $-6.81^{+0.10}_{-5.30}$ & $-4.53^{+0.26}_{-4.79}$ & $-1.86^{+0.07}_{-0.04}$ & $-1.00^{+0.06}_{-0.11}$ & $0.113^{+0.015}_{-0.020}$ & $4.33$ \\
    \gtwentyone \, MgSiO$_3$ & $-6.62^{+0.06}_{-5.30}$ & $-4.42^{+0.00}_{-4.68}$ & $-1.86^{+0.12}_{-0.13}$ & $-0.57^{+0.06}_{-0.08}$ & $0.284^{+0.039}_{-0.040}$ & $3.34$ \\
    \hline
    \gtwentyone \, ACAR (no $160 \um$) & $-6.79^{+0.10}_{-5.28}$ & $-4.09^{+0.39}_{-5.23}$ & $-1.72^{+0.06}_{-0.00}$ & $-1.89^{+0.12}_{-2.03}$ & $0.032^{+0.005}_{-0.011}$ & $1.32$ \\
    \gtwentyone \, BE (no $160 \um$) & $-6.79^{+0.04}_{-5.33}$ & $-4.49^{+0.22}_{-4.85}$ & $-1.78^{+0.08}_{-0.03}$ & $-1.41^{+0.21}_{-1.99}$ & $0.055^{+0.023}_{-0.035}$ & $1.19$ \\
    \gtwentyone \, MgSiO$_3$ (no $160 \um$) & $-6.63^{+0.05}_{-5.27}$ & $-4.59^{+0.03}_{-4.50}$ & $-1.65^{+0.13}_{-0.13}$ & $-0.88^{+0.22}_{-0.91}$ & $0.156^{+0.080}_{-0.108}$ & $0.74$ \\
    \hline
    \gtwentynine \, ACAR & $-4.91^{+0.62}_{-6.75}$ & $-2.63^{+0.15}_{-3.92}$ & $-2.15^{+0.45}_{-3.71}$ & $-2.09^{+0.30}_{-1.79}$ & $0.018^{+0.013}_{-0.013}$ & $1.66$ \\
    \gtwentynine \, BE & $-4.94^{+0.61}_{-6.73}$ & $-2.78^{+0.14}_{-1.83}$ & $-1.83^{+0.22}_{-0.96}$ & $-1.82^{+0.33}_{-2.00}$ & $0.031^{+0.015}_{-0.017}$ & $1.30$ \\
    \gtwentynine \, MgSiO$_3$ & $-4.60^{+0.48}_{-6.69}$ & $-2.51^{+0.21}_{-4.46}$ & $-1.58^{+0.28}_{-2.29}$ & $-1.31^{+0.38}_{-2.14}$ & $0.079^{+0.057}_{-0.049}$ & $0.96$ \\
    \hline
  \end{tabular}
  \label{tab:pwndust}
\end{table*}

\begin{figure*}
  \centering
  \subfigure{\includegraphics[width=\columnwidth]{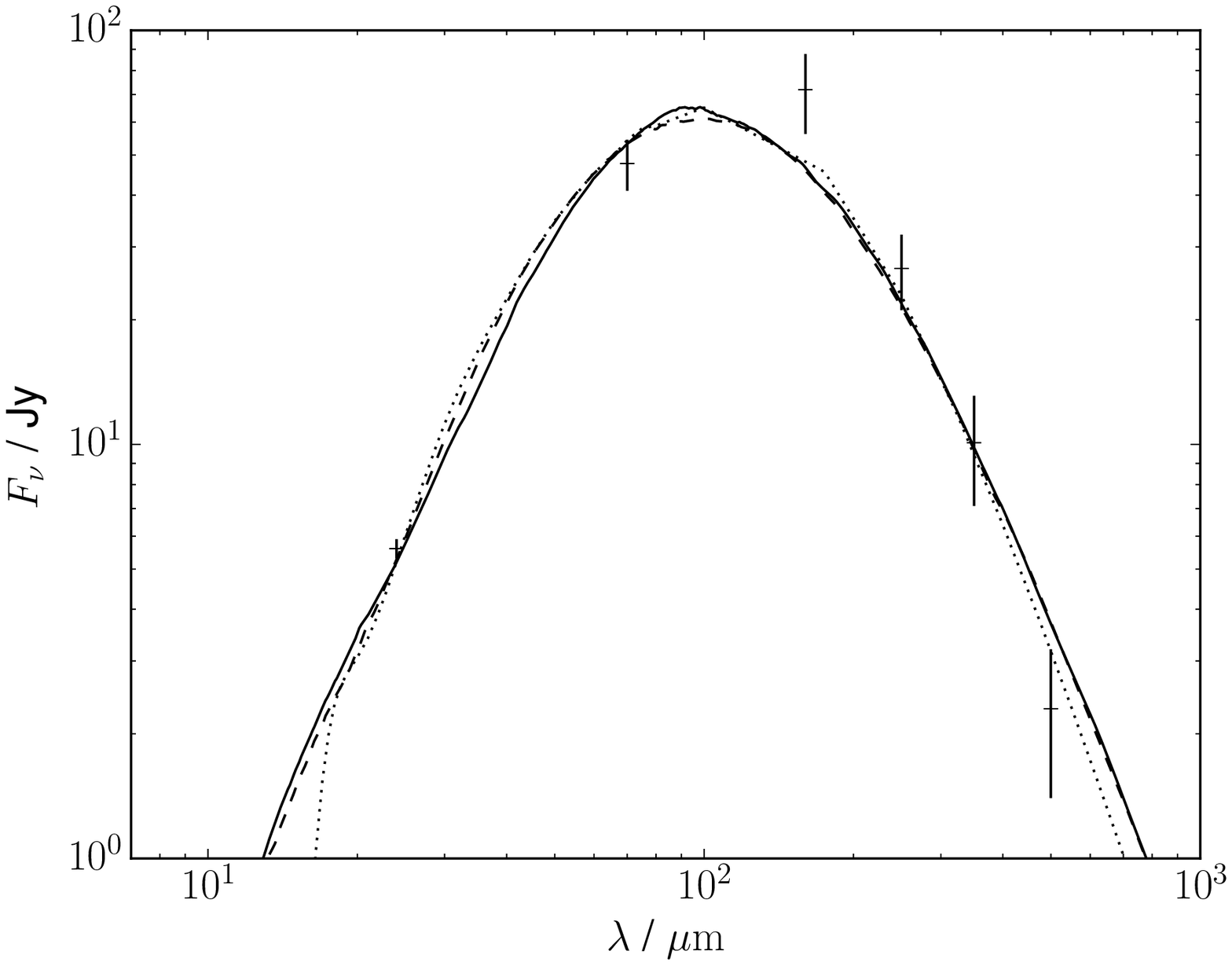}}\quad
  \subfigure{\includegraphics[width=\columnwidth]{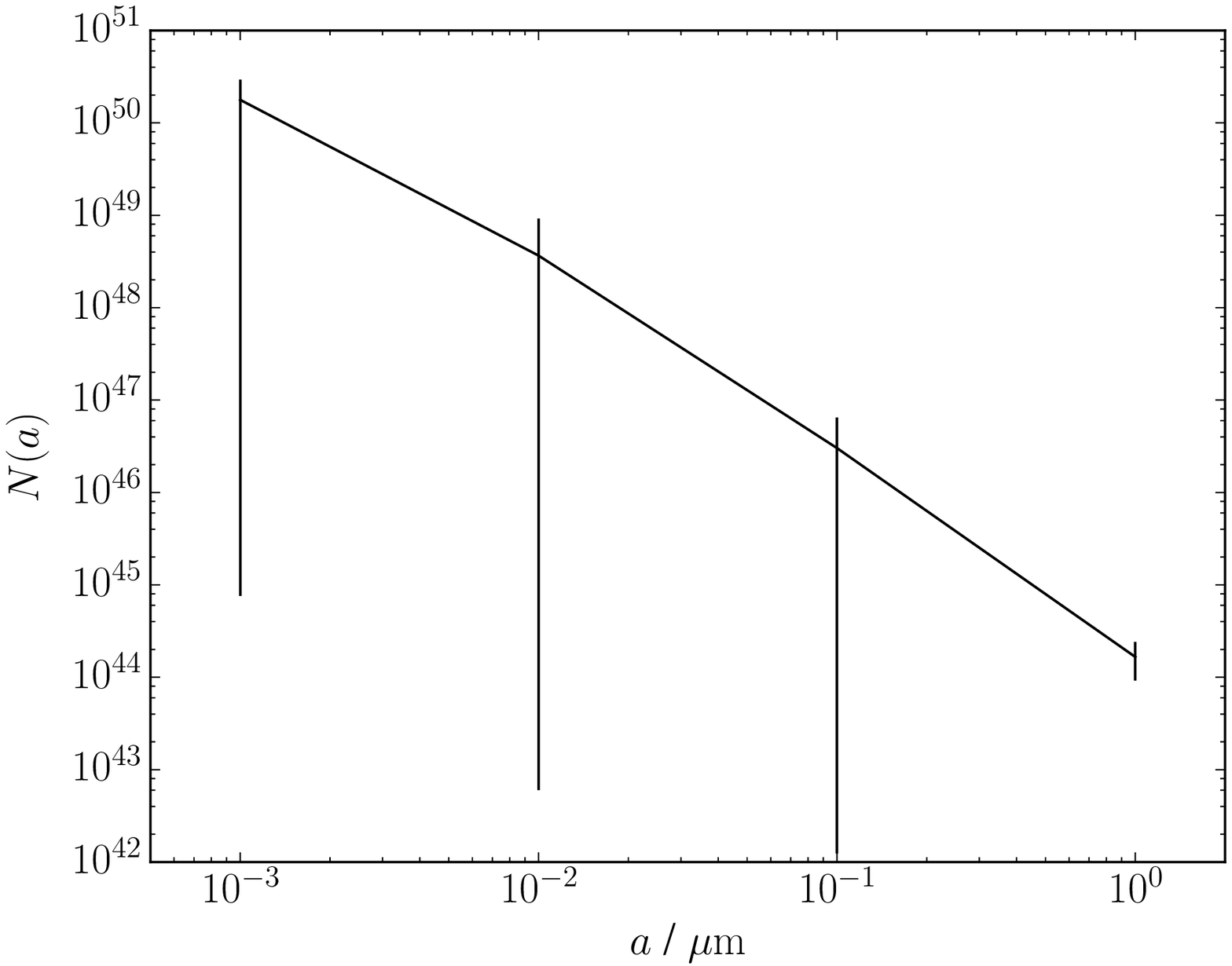}}
  \caption{Left: Model SEDs for \geleven \, using ACAR (solid line), BE (dashed line) and MgSiO$_3$ (dotted line) grains and SNR dust fluxes from \citet{chawner2019} (black crosses). Right: Grain size distribution for \geleven \, using BE grains.}
  \label{fig:g11}
\end{figure*}

Figure \ref{fig:g11} shows the total dust SEDs and the size distribution for BE grains for \geleven. \geleven \, is a significantly worse fit than the other two PWN, with $\chi^2 > 5$. There is a noticeable discrepancy between the $160 \um$ and, to a lesser extent, the $250 \um$ fluxes, where our models underpredict the observed values. \citet{chawner2019}, in a point-process mapping (PPMAP; \citealt{marsh2015}) analysis of \geleven, found that compared to their background-subtracted fluxes PPMAP returned lower far-IR values for the dust emission associated with the SNR, particularly the $160 \um$ flux. If the values we use are significantly contaminated by ISM dust emission, the somewhat poor fit in this region may be an indication that these fluxes cannot be (entirely) produced by PWN-heated ejecta dust. { The $160 \um$ flux may also be contaminated by [C II] emission, as in \gtwentynine \, \citep{temim2019}.} Our carbon grain models require dust masses comparable to the value of $0.34 \pm 0.14 \msun$ from \citet{chawner2019} (using $\kappa_{850 \um} = 0.7 \, {\rm cm^2 g^{-1}}$) - the MgSiO$_3$ dust mass is $> 1 \msun$, which would require a very large (although not necessarily implausible, e.g. \citealt{woosley1995}) mass of metals in the ejecta. The grain size distribution is approximately a power law, although with large error bars on all but the $1.0 \um$ grains. Unlike the Crab Nebula, the grain size distribution does not flatten for small radii, so there is a substantial contribution to the SED from small grains, particularly in the mid-IR.

\begin{figure*}
  \centering
  \subfigure{\includegraphics[width=\columnwidth]{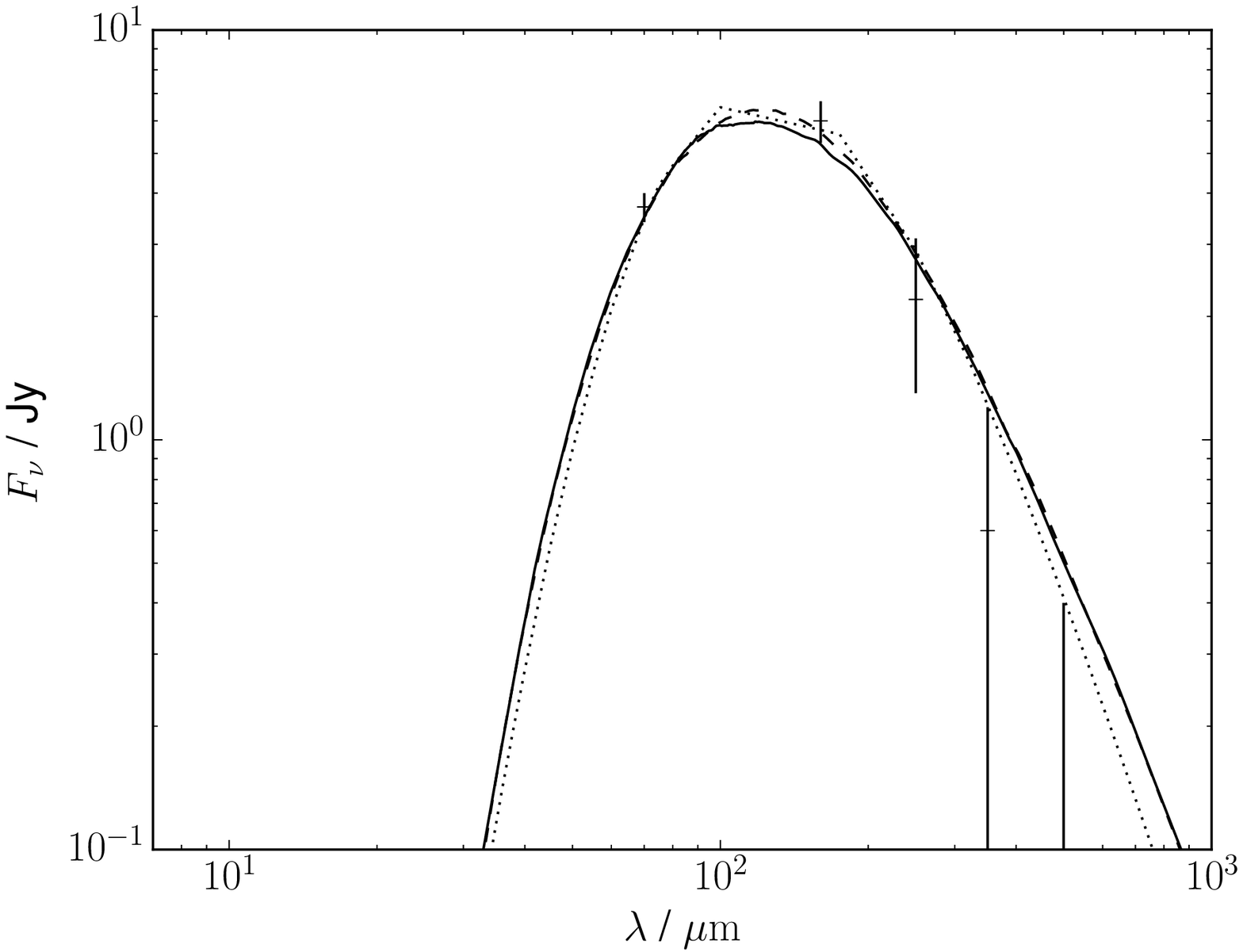}}\quad
  \subfigure{\includegraphics[width=\columnwidth]{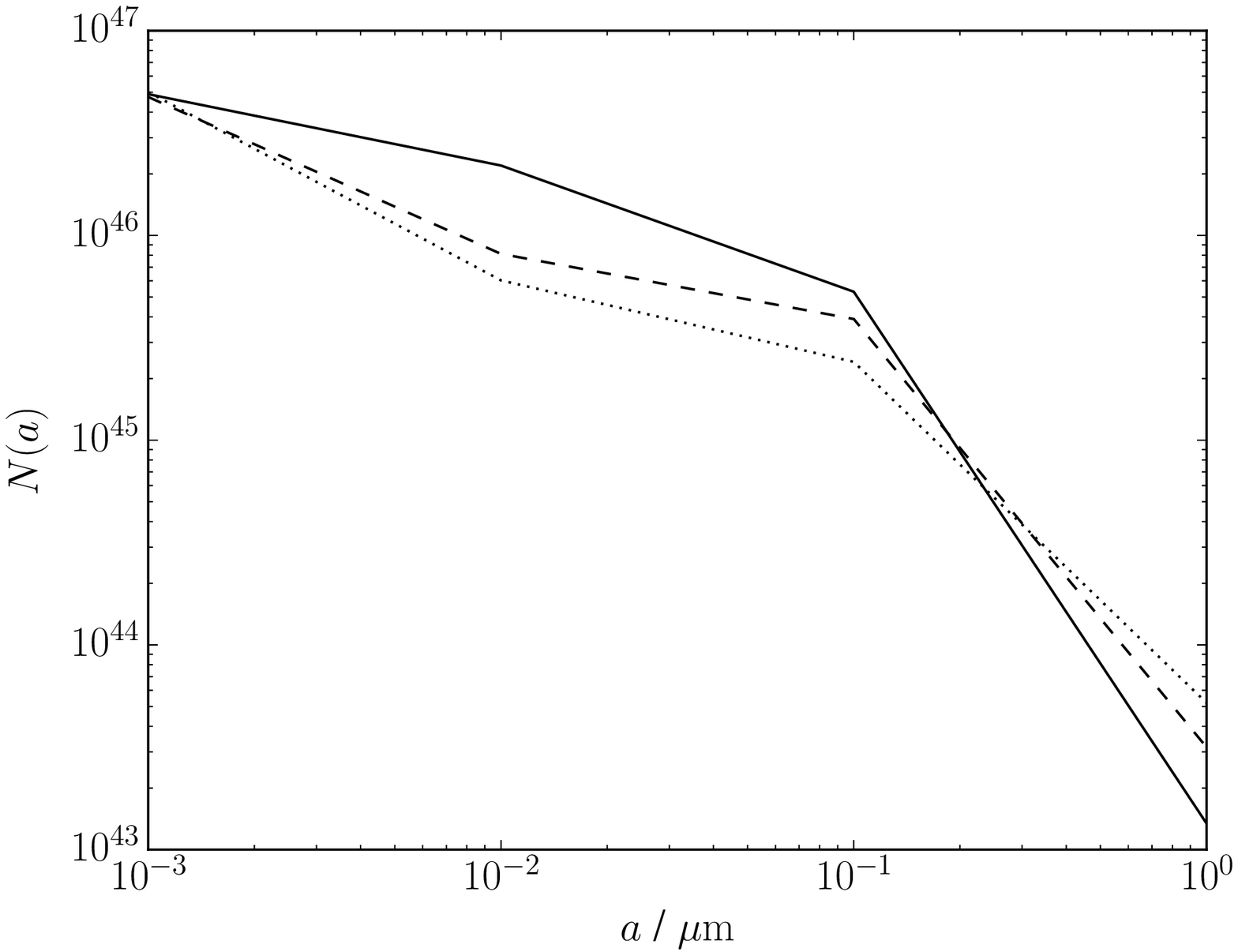}}
  \caption{Left: Model SEDs for \gtwentyone \, using ACAR (solid line), BE (dashed line) and MgSiO$_3$ (dotted line) grains and SNR dust fluxes from \citet{chawner2019} (black crosses). Right: Grain size distributions for \gtwentyone \, using ACAR (solid line), BE (dashed line) and MgSiO$_3$ (dotted line) grains.}
  \label{fig:g21}
\end{figure*}

Figure \ref{fig:g21} shows the dust SEDs and grain size distributions for \gtwentyone \, - for clarity, we do not show the error bars on $\log N(a)$. Again, the SEDs for different grain species are similar, although in this case BE and MgSiO$_3$ grains are somewhat better fits than ACAR grains. The MgSiO$_3$ dust mass, $0.28 \pm 0.04 \msun$, is in very good agreement with the PPMAP value from \citet{chawner2019} ($0.29 \pm 0.08 \msun$). The grain size distribution is similar to the Crab Nebula, although changes slope at a larger radius, and so micron-sized grains comprise an even larger fraction of the total dust mass. Our models slightly overestimate the far-IR flux, the opposite of the situation in \geleven. However, in this case \citet{chawner2019} find higher far-IR fluxes than the background subtracted values we use from PPMAP analysis - the use of a more detailed method to separate remnant and ISM emission again reduces the discrepancy between our models and observations. { Alternatively, if the $160 \um$ flux is contaminated by line emission, the best-fit model may be producing too much far-IR flux in an attempt to match this unrealistically high value. \citet{temim2017b} report potentially significant [C II] $157 \um$ emission from this PWN, although no line correction factor is available. Repeating the modelling without this data point, the fit is significantly improved, with dust masses reduced by a factor of $\sim 2$ so that the range is now $0.02-0.24 \msun$, with the grain size distribution not substantially affected.}

\begin{figure*}
  \centering
  \subfigure{\includegraphics[width=\columnwidth]{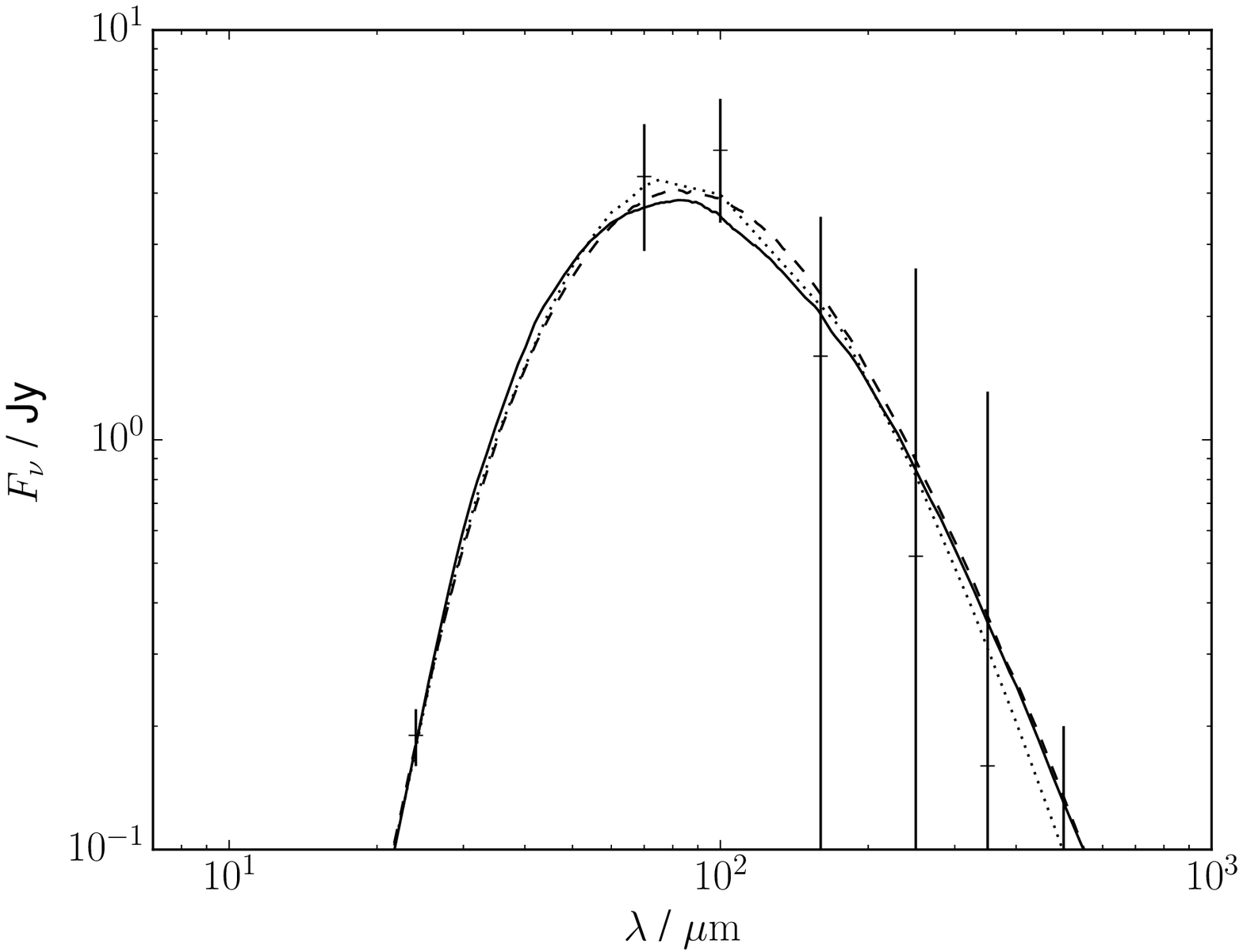}}\quad
  \subfigure{\includegraphics[width=\columnwidth]{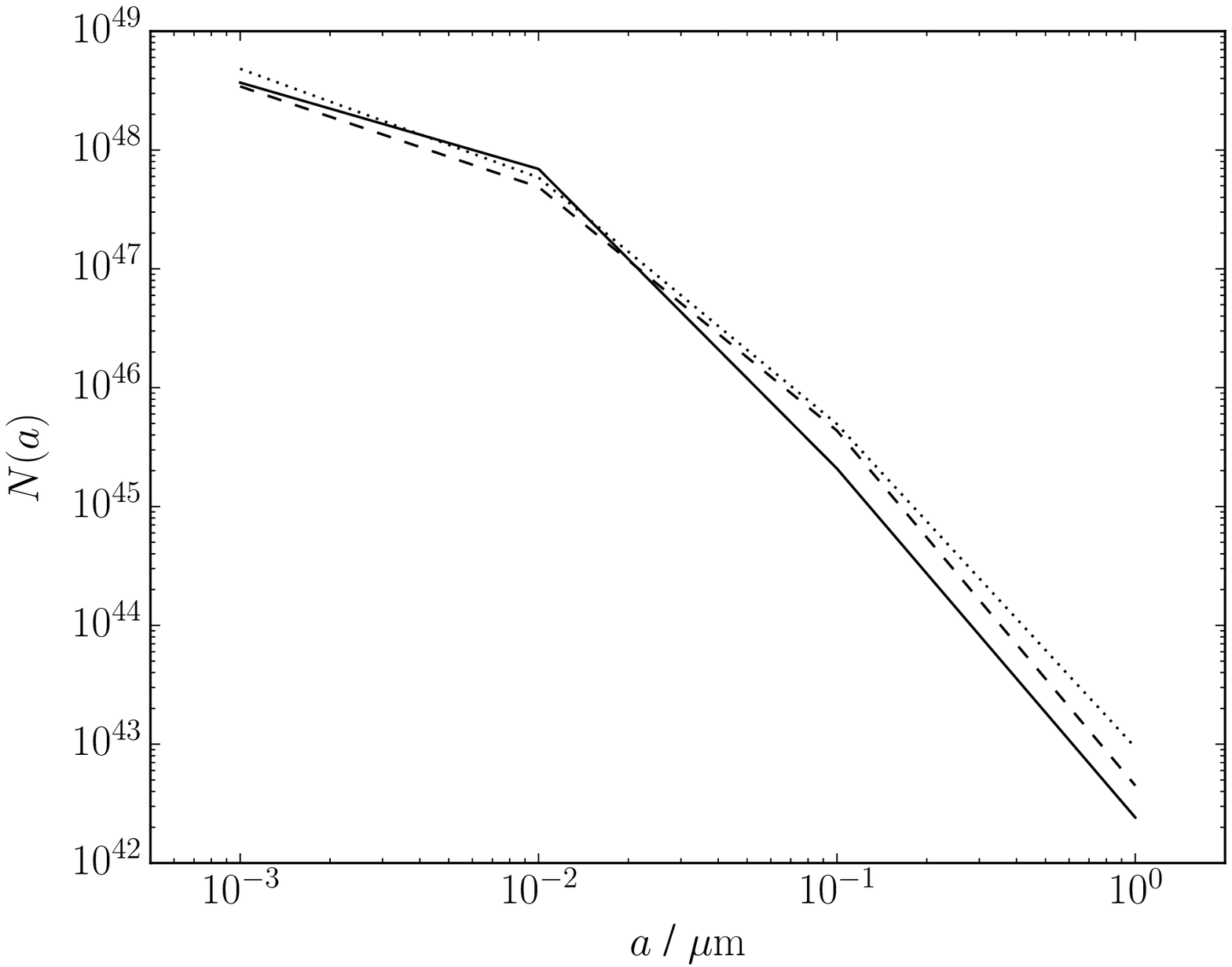}}
  \caption{Left: Model SEDs for \gtwentynine \, using ACAR (solid line), BE (dashed line) and MgSiO$_3$ (dotted line) grains and SNR dust fluxes from \citet{chawner2019} { and \citet{temim2019}} (black crosses). Right: Grain size distributions for \gtwentynine \, using ACAR (solid line), BE (dashed line) and MgSiO$_3$ (dotted line) grains.}
  \label{fig:g29}
\end{figure*}

Figure \ref{fig:g29} shows the dust SEDs and grain size distributions for \gtwentynine \, (also known as Kes 75). For this object, the uncertainties on the synchrotron-subtracted far-IR fluxes are larger than the absolute values, meaning that the relative uncertainties on the total dust mass are much higher than for the other PWNe. However, we still find minimum masses $>0.01 \msun$ for all grain species { (although ACAR grains are consistent with a smaller dust mass)}. Unsurprisingly given the large error bars, this is formally our best-fit SNR, with MgSiO$_3$ grains again marginally preferred. The grain size distributions appear intermediate between the Crab Nebula and \gtwentyone. { \citet{temim2019} found dust masses of $0.003-0.08 \msun$ based on single-temperature fits to the PACS data, consistent with our values although our mass range is shifted to slightly higher values. \citet{temim2019} regarded the $24 \um$ emission as unrelated to the PWN dust, using an upper limit of $0.2 \, {\rm Jy}$ - however, we find that our results are insensitive to the $24 \um$ flux, with very little change even for a stricter upper limit of $0.03 \, {\rm Jy}$ (the uncertainty on the measurement from \citet{chawner2019}). Our inferred dust masses are much lower than the PPMAP value of $0.51 \pm 0.13 \msun$ from \citet{chawner2019} - however, these authors used a distance of $10.6 \kpc$, around twice our adopted value, leading to a factor of $\sim 4$ difference in the luminosities (and therefore dust masses). Accounting for this, our silicate dust mass is consistent with their value.}

\subsubsection{Grain heating mechanisms}

\citet{temim2019} suggest that the dust in \gtwentynine \, must be shock heated, as the PWN radiation field is not sufficient to heat even very small grains to the temperatures required by their blackbody fits. { In fact, we find that the synchrotron radiation is insufficient to produce the observed dust emission for each of these three PWNe, failing to fit the observed $24$ and $70 \um$ fluxes. Our models' ability to fit the data is due to the inclusion of collisional heating by electrons, for which we took the density and temperature of the Crab Nebula. These parameters are not necessarily appropriate for the other PWNe, and in the case of \gtwentynine \, heating by shocked gas may be more realistic. We investigate two additional scenarios for the collisional heating: photoionized gas, with $\nel = 10 \pcc$ and $\tel = 10^4 \kel$, and shock-heated gas, with $\nel = 100 \pcc$ and $\tel = 10^6 \kel$ as in the shock model used by \citet{temim2019} for the line emission. The photoionized case fails to heat grains of any size to the necessary temperature, as the density is too low. The shocked models also fail for \geleven \, and \gtwentyone, as even the largest grains are at too high a temperature to fit the far-IR data without exceeding the shorter wavelength fluxes. Dust masses for \gtwentynine \, in the shocked case are listed in Table \ref{tab:shockdust} - the models are formally a better fit than our original ones, although given the uncertainties involved we do not consider this conclusive evidence. Dust masses are reduced in all cases, as the dust temperatures are higher, which also results in severe limits on the mass present in all but the largest grain sizes - the mass fraction in micron-sized grains is essentially unity. The carbon dust masses are in good agreement with the value from \citet{temim2019}, while the silicate value is lower, due to a higher grain temperature ($43 \kel$ for $1.0 \um$ grains, compared to $33 \kel$ in \citet{temim2019}).}

\begin{table*}
  \centering
  \caption{{ Mass of dust grains of different sizes, total dust masses and $\chi^2$ values for different grain species in \gtwentynine, with $\nel = 100 \pcc$ and $\tel = 10^6 \kel$.}}
  \begin{tabular}{c|cccc|cc}
    \hline
    & \multicolumn{4}{|c|}{$\log M_{\rm dust}(a)$/$\msun$} &  \\
    Model & $0.001 \um$ & $0.01 \um$ & $0.1 \um$ & $1.0 \um$ & $M_{\rm tot}$/$\msun$ & $\chi^2$ \\
    \hline
    \gtwentynine \, ACAR & $-5.42^{+0.43}_{-6.23}$ & $-5.92^{+0.43}_{-3.25}$ & $-4.40^{+0.32}_{-1.99}$ & $-2.16^{+0.07}_{-0.09}$ & $0.007^{+0.001}_{-0.001}$ & $0.65$ \\
    \gtwentynine \, BE & $-5.93^{+0.00}_{-6.02}$ & $-6.41^{+0.33}_{-3.09}$ & $-4.98^{+0.36}_{-1.88}$ & $-2.10^{+0.06}_{-0.07}$ & $0.008^{+0.001}_{-0.001}$ & $0.82$ \\
    \gtwentynine \, MgSiO$_3$ & $-6.94^{+0.06}_{-5.07}$ & $-6.58^{+0.23}_{-2.94}$ & $-4.98^{+0.30}_{-1.77}$ & $-1.91^{+0.05}_{-0.06}$ & $0.012^{+0.002}_{-0.002}$ & $1.08$ \\
    \hline
  \end{tabular}
  \label{tab:shockdust}
\end{table*}

{ While different parameters for \geleven \, and \gtwentyone \, may allow a `shocked' model to successfully fit the data for these objects, we consider our initial models to be reasonable. The warm, dense gas in the Crab Nebula is the result of heating by the charged particle flux from the PWN \citep{richardson2013,priestley2017}, which presumably also occurs in the other PWNe. If the dust is, in fact, shock heated in these objects, any reasonable model would produce similar results as for \gtwentynine, so our derived dust masses would be smaller with a higher fraction of the total mass in large grains. For \gtwentynine, line observations indicate that there is interaction between the PWN and the ejecta \citep{temim2019}, so the shock model may be better motivated. However, we note that the model adopted by \citet{temim2019} results in a downstream photoionization region with densities and temperatures very similar to those adopted in our initial model, and a much greater proportion of the mass than the $10^6 \kel$ shocked region, so dust heating by warm gas is not unrealistic. The dust-to-gas mass ratios in this case (based on a swept-up ejecta mass of $0.1 \msun$; \citealt{temim2019}) are high ($\sim 0.2$ for carbon, $\sim 0.8$ for silicates), but not necessarily unreasonable considering the values for Cas A and the Crab Nebula are $\gtrsim 0.2$ \citep{priestley2019} and $\sim 0.1$ (\citealt{delooze2019}, this work) respectively.} { If the warm gas is charged particle heated material, as in the Crab Nebula, rather than post-shock material, the ejecta gas mass corresponding to the dust would be larger than the $0.1 \msun$ currently swept up by the expanding PWNe, and the dust-to-gas mass ratio lower.}

\subsection{\gfiftyfour}

\begin{figure*}
  \centering
  \subfigure{\includegraphics[width=\columnwidth]{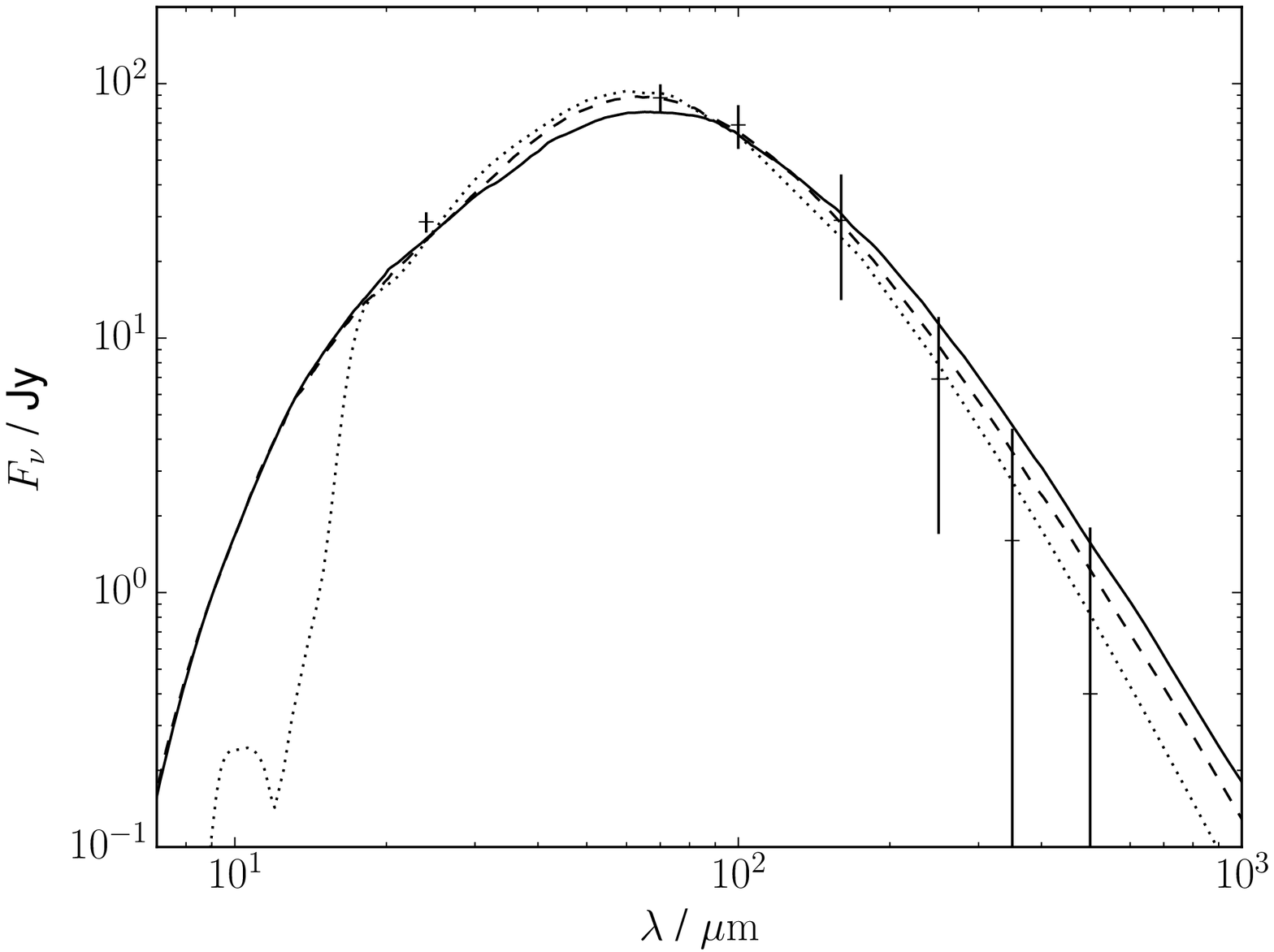}}\quad
  \subfigure{\includegraphics[width=\columnwidth]{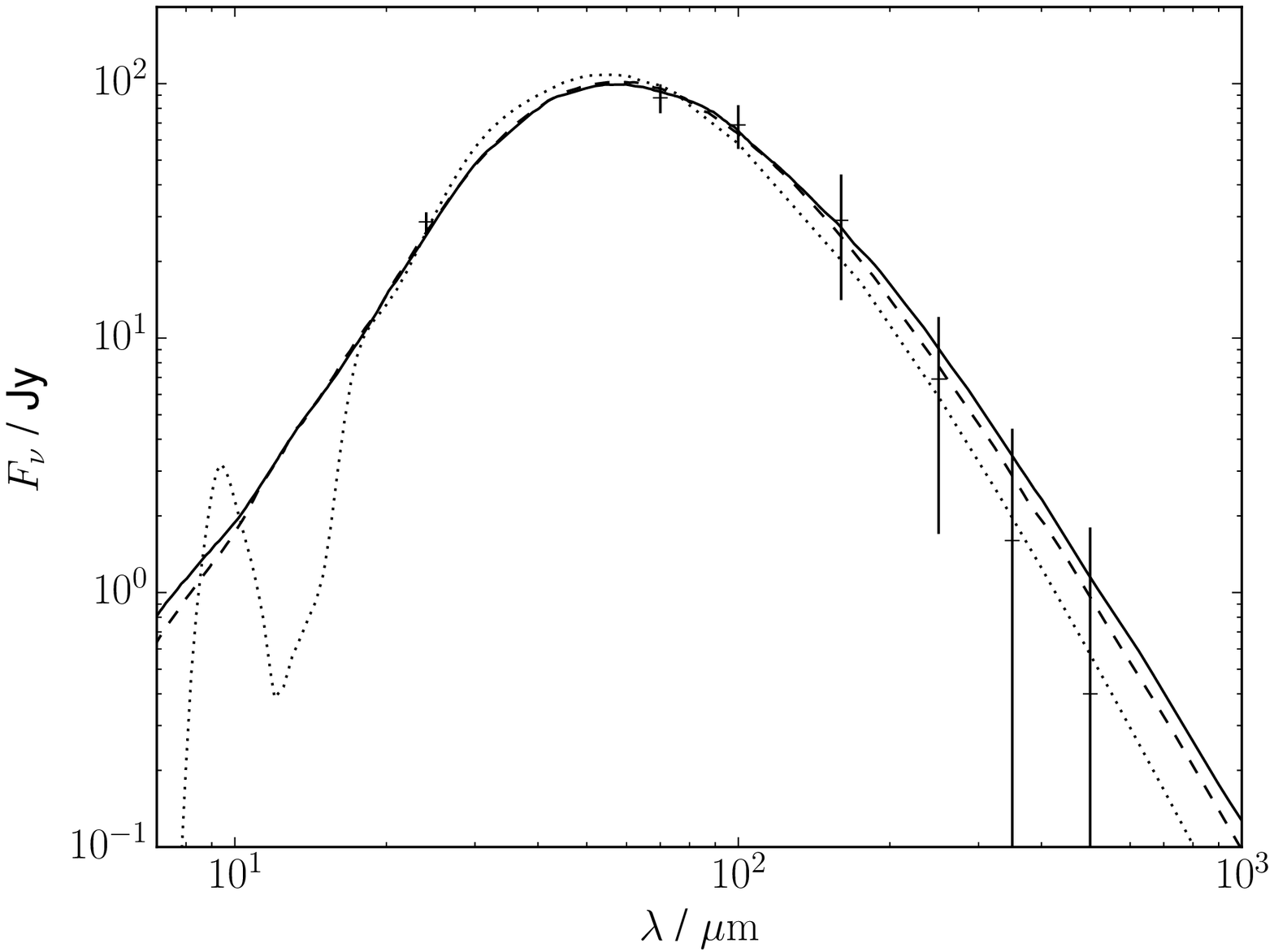}}
  \caption{Model SEDs for \gfiftyfour \, using ACAR (solid line), BE (dashed line) and MgSiO$_3$ (dotted line) grains, heated by the PWN synchrotron radiation (left) and OB stars { at $2 \pc$} (right), and SNR dust fluxes from \citet{temim2017} (black crosses).}
  \label{fig:g54flux}
\end{figure*}

Figure \ref{fig:g54flux} shows the dust SEDs for \gfiftyfour \, for { heating by PWN synchrotron radiation and nearby OB stars at a distance of $2 \pc$.} The required dust masses are listed in Table \ref{tab:g54dust}. We find { total dust masses of $0.08-0.91 \msun$ (PWN), $0.05-0.45 \msun$ (OB $2 \pc$) and $0.04-0.1 \msun$ (OB $0.2 \pc$)}, with silicates again providing both the highest dust masses and the best fits to the data { except for the $0.2 \pc$ OB star model.} This is unsurprising, as the mid-IR spectrum of \gfiftyfour \, contains features identified with magnesium silicate grains, both in this SNR and in Cassiopeia A \citep{rho2008,temim2017}. We also find that the OB models are better fits to the data than the PWN ones, as the predicted dust temperatures are higher and the models are better able to fit the $24 \um$ point, which is the main discrepancy. \citet{rho2018}, using carbon grains as their cool dust component, found a total mass (the majority of which is carbon) of $0.26 \pm 0.05 \msun$, consistent with our ACAR and BE PWN models but significantly larger than our OB carbon dust masses, even assuming the maximum allowed values. They also investigated various silicate grains - their preferred composition (Mg$_2$SiO$_4$) results in a total dust mass of $0.9 \pm 0.3 \msun$, again consistent within the error bars with our PWN model but higher than the { OB MgSiO$_3$ masses}. Their values for other silicate compositions are more similar to our { OB models}, although they disfavour these compositions.

\begin{figure}
  \centering
  \includegraphics[width=\columnwidth]{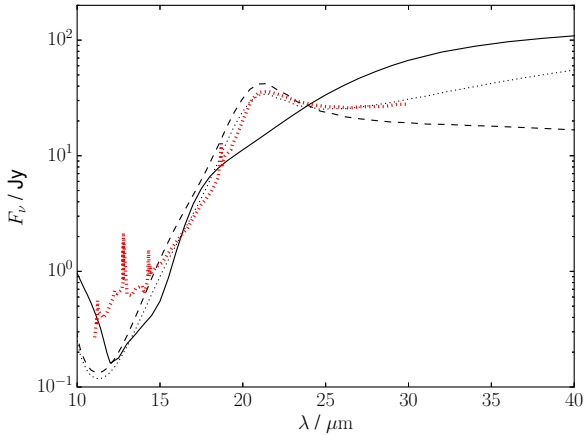}
  \caption{Model SEDs for \gfiftyfour \, using the OB $0.2 \pc$ models for MgSiO$_3$ (black solid line) and Mg$_{0.7}$SiO$_{2.7}$ (black dashed line) and the OB $2 \pc$ model for Mg$_{0.7}$SiO$_{2.7}$ (black dotted line), and the Spitzer IRS spectrum from position 1 of \citet{temim2010} (red stippled line) scaled to the $24 \um$ flux.}
  \label{fig:21um}
\end{figure}

\citet{temim2017} found dust masses { of at least $0.26 \msun$ for \gfiftyfour \, assuming that Mg$_{0.7}$SiO$_{2.7}$ grains are responsible for much of the emission, in good agreement with our silicate masses for the PWN and $2 \pc$ OB models. Our $0.2 \pc$ OB silicate masses are constrained to be less than half this value, due to both a minimum grain temperature ($55 \kel$) above the largest values found by \citet{temim2019} (excluding their hot component, which contributes negligibly to the mass), and the lower emissivity per unit mass of their silicate composition (e.g. \citealt{delooze2017}) compared to MgSiO$_3$. Using Mg$_{0.7}$SiO$_{2.7}$ optical constants from \citet{jaeger2003} rather than MgSiO$_3$, we find an improved fit for the $2 \pc$ OB model, { but this requires $3.4^{+0.6}_{-0.5} \msun$ of dust}, again mostly in micron-sized grains. The PWN model has a similar best-fit dust mass, but with much larger error bars, as in this case the constraints on smaller grain sizes are not as severe. For a distance of $0.2 \pc$, the ratio of $24 \um$ to far-IR flux is too high to fit the data. Our best-fit model in this case { requires $0.032^{+0.004}_{-0.002} \msun$ of $1.0 \um$ grains} to fit the $24 \um$ flux while significantly underpredicting the values at the other wavelengths - using a different species to produce the far-IR emission in combination with Mg$_{0.7}$SiO$_{2.7}$ grains, as in \citet{temim2017}, could resolve this issue.} { Figure \ref{fig:21um} shows the mid-IR spectrum from \citet{temim2010}, scaled to the total $24 \um$ flux, and three silicate models. The MgSiO$_3$ model clearly fails to reprodue the observed spectral features, whereas the two Mg$_{0.7}$SiO$_{2.7}$ OB models are consistent with the data, although the $0.2 \pc$ model does not fit the far-IR data and the $2 \pc$ model would require an extremely high dust mass to do so, suggesting multiple dust species are required.}

\begin{table*}
  \centering
  \caption{{ Mass of dust grains of different sizes, total dust masses and $\chi^2$ values for different grain species and heating sources for \gfiftyfour.}}
  \begin{tabular}{c|cccc|cc}
    \hline
    & \multicolumn{4}{|c|}{$\log M_{\rm dust}(a)$/$\msun$} &  \\
    Model & $0.001 \um$ & $0.01 \um$ & $0.1 \um$ & $1.0 \um$ & $M_{\rm tot}$/$\msun$ & $\chi^2$ \\
    \hline
    PWN ACAR & $-2.11^{+0.12}_{-2.92}$ & $-1.28^{+0.17}_{-6.51}$ & $-1.06^{+0.56}_{-4.98}$ & $-1.54^{+0.06}_{-2.28}$ & $0.176^{+0.169}_{-0.096}$ & $6.21$ \\
    PWN BE & $-2.16^{+0.14}_{-5.28}$ & $-1.17^{+0.16}_{-5.37}$ & $-1.00^{+0.60}_{-5.07}$ & $-1.23^{+0.07}_{-2.51}$ & $0.233^{+0.221}_{-0.138}$ & $4.26$ \\
    PWN MgSiO$_3$ & $-2.01^{+0.15}_{-5.16}$ & $-0.95^{+0.18}_{-4.78}$ & $-0.64^{+0.53}_{-4.99}$ & $-0.69^{+0.16}_{-2.73}$ & $0.555^{+0.358}_{-0.378}$ & $3.58$ \\
    { PWN Mg$_{0.7}$SiO$_{2.7}$} & $-2.76^{+0.39}_{-7.94}$ & $-1.40^{+0.32}_{-5.83}$ & $0.15^{+0.16}_{-3.20}$ & $0.40^{+0.51}_{-3.36}$ & $3.967^{+4.229}_{-2.030}$ & $0.39$ \\
    \hline
    OB $2 \pc$ ACAR & $-2.40^{+0.32}_{-6.47}$ & $-1.51^{+0.21}_{-4.78}$ & $-1.50^{+0.53}_{-4.48}$ & $-1.49^{+0.08}_{-2.29}$ & $0.098^{+0.038}_{-0.050}$ & $2.58$ \\
    OB $2 \pc$ BE & $-2.42^{+0.34}_{-5.87}$ & $-1.46^{+0.25}_{-5.63}$ & $-1.35^{+0.47}_{-4.73}$ & $-1.18^{+0.16}_{-2.40}$ & $0.150^{+0.041}_{-0.087}$ & $1.97$ \\
    OB $2 \pc$ MgSiO$_3$ & $-2.46^{+0.52}_{-8.34}$ & $-1.23^{+0.20}_{-5.47}$ & $-1.10^{+0.51}_{-4.82}$ & $-0.76^{+0.23}_{-2.56}$ & $0.315^{+0.130}_{-0.219}$ & $1.76$ \\
    { OB $2 \pc$ Mg$_{0.7}$SiO$_{2.7}$} & $-3.10^{+0.14}_{-8.19}$ & $-2.22^{+0.54}_{-6.13}$ & $-0.75^{+0.23}_{-3.50}$ & $0.50^{+0.09}_{-0.09}$ & $3.378^{+0.576}_{-0.480}$ & $0.26$ \\
    \hline
    OB $0.2 \pc$ ACAR & $-4.22^{+0.34}_{-7.31}$ & $-4.37^{+0.43}_{-4.30}$ & $-3.52^{+0.32}_{-2.41}$ & $-1.38^{+0.04}_{-0.04}$ & $0.042^{+0.003}_{-0.004}$ & $1.39$ \\
    OB $0.2 \pc$ BE & $-4.79^{+0.22}_{-6.95}$ & $-5.07^{+0.12}_{-4.29}$ & $-4.19^{+0.24}_{-2.54}$ & $-1.32^{+0.03}_{-0.02}$ & $0.048^{+0.004}_{-0.003}$ & $2.55$ \\
    OB $0.2 \pc$ MgSiO$_3$ & $-4.95^{+0.15}_{-6.68}$ & $-5.06^{+0.29}_{-4.01}$ & $-3.80^{+0.30}_{-2.61}$ & $-1.03^{+0.04}_{-0.03}$ & $0.093^{+0.008}_{-0.006}$ & $2.93$ \\
    { OB $0.2 \pc$ Mg$_{0.7}$SiO$_{2.7}$} & $-5.20^{+0.53}_{-6.54}$ & $-5.48^{+0.03}_{-3.86}$ & $-3.90^{+0.10}_{-2.72}$ & $-1.49^{+0.05}_{-0.02}$ & $0.032^{+0.004}_{-0.002}$ & $79.58$ \\
    \hline
  \end{tabular}
  \label{tab:g54dust}
\end{table*}

Continuing the trend seen with the previous SNRs in this paper, we find that \gfiftyfour \, must contain significant masses of micron-sized dust grains. For the PWN models, there is slightly more mass in the $0.1 \um$ grains, whereas the OB models have more mass at larger grain size, { particularly for the $0.2 \pc$ case which, like the shock models for \gtwentynine, requires negligible quantities of grains below micron-sized}. Unlike the previous objects we find a relatively large ($\gtrsim 10^{-3} \msun$) mass of $0.001 \um$ grains, along with significant masses (up to $\sim 0.1 \msun$) of $0.01 \um$ grains, { for the PWN and $2 \pc$ OB models} - \gfiftyfour \, has by far the largest $24 \um$ flux, so requires more small grains which are { at high enough temperatures} to emit strongly at these wavelengths. { For Mg$_{0.7}$SiO$_{2.7}$ grains, which have an emission feature at $21 \um$, the masses of small grains are lower, although still larger than in the other PWNe.} The grain size distributions, similarly to \geleven, are close to power laws in shape. { For the $0.2 \pc$ OB models, large grains are heated strongly enough to emit significantly at shorter wavelengths, and smaller grains can only be present in very small quantities.}

{ As with \geleven, \gtwentyone \, and \gtwentynine, the synchrotron luminosity in our PWN model is insufficient to power the observed dust emission, and additional collisional heating is required to fit the SED. The luminosities of our OB heating models, even for an extreme distance of $2 \pc$, are large enough to heat grains to sufficiently high temperatures to fit the $24 \um$ flux without invoking any additional heating sources. Additionally, both OB models give lower $\chi^2$ values, and as such we regard this scenario as more plausible than the synchrotron-heated case suggested by \citet{rho2018}.} { Our OB models, which assume all the dust is located at a single distance from a single star, is clearly unphysical, as even for a single star we would expect dust to be distributed over a range of radii, such as the model used by \citet{temim2010}, and the true geometry of the object is far more complicated. However, given that \citet{temim2010} found that the dust is likely optically thin, we can at least constrain the possible dust properties - it is unlikely that the maximum distance from a star is greater than $2 \pc$, so this model gives a rough upper limit on the allowed dust mass for a given composition. At a distance of $0.2 \pc$ we already find that the mass fraction in micron-sized grains is essentially unity, and closer distances (down to $0.003 \pc$ in the \citet{temim2010} model) can only realistically increase the fraction of large grains. The required mass in this situation would be lower, and our model is unable to provide a lower bound, but by the same logic the mass fraction in micron-sized grains can be constrained to be at least the $\sim 50 \%$ found for the $2 \pc$ models.}

\section{Discussion \& Conclusions}

The five SNRs considered in this paper have ages ranging { from $\sim 500-2000 \yr$ \citep{bocchino2010,reynolds2018}}, and include the extremes of possible CCSNe progenitor systems - the Crab Nebula is assumed to have been a type IIP explosion of a low-mass ($\sim 8 \msun$) progenitor \citep{smith2013}, { while \citet{borkowski2016} suggested \geleven \, originated from a stripped-envelope CCSNe, implying an initial stellar mass $\gtrsim 20 \msun$.} While the Crab Nebula has no detectable forward or reverse shock, \geleven \, and \gtwentynine \, show strong interactions with the surrounding ISM, and \citet{borkowski2016} have claimed that \geleven \, has already been fully swept by the reverse shock. Despite these differences, the dust { size distributions we find show remarkable similarities. In all cases we find that $0.1$ and $1.0 \um$ grains make up virtually all the dust mass, while only \geleven \, and \gfiftyfour \, show any evidence of grains $< 0.01 \um$ in size, and our models of \gtwentyone \, do not even require $0.01 \um$ grains. Most ISM-type dust models also require increasing mass fractions in larger grain sizes, but for an MRN-like size distribution with an exponent of $-3.5$, the mass between $0.5$ and $1.0 \um$ is $\sim 30 \%$, whereas we only find comparable values for our (disfavoured) PWN models of \gfiftyfour \, - excluding these models, the mass fraction in micron-sized grains is never below $50 \%$ and often significantly greater. While in several cases the heating mechanism responsible for the dust emission is unclear, and can have significant effects on our derived dust masses, the alternative cases we have examined only strengthen the evidence for micron-sized grains. Combined with previous studies of other nearby SNRs \citep{owen2015,wesson2015,bevan2016}, this suggests that micron-sized dust grains may be ubiquitous in SNR ejecta dust.}

{ While the presence of micron-sized grains seems to be robust to the choice of grain heating mechanism, this can cause our dust masses to vary considerably, along with the choice of grain composition. For the Crab Nebula, while the adopted distance from the PWN causes some variation, this is well within a factor of a few, and a value of $\sim 0.05 \msun$ of carbon dust in agreement with \citet{delooze2019} seems relatively secure. For the three objects requiring collisional heating (\geleven, \gtwentyone \, and \gtwentynine), our values listed in Table \ref{tab:pwndust} are most likely an upper bound, as realistic alternative heating processes are likely to involve significantly higher gas temperatures. As seen in Table \ref{tab:shockdust}, this can result in an inferred dust mass lower by a factor of $\gtrsim 5$ than for the Crab Nebula collisional heating parameters, although we note that the PPMAP analysis of \citet{chawner2019} required cold dust masses comparable to our initial values (i.e. $\gtrsim 0.1 \msun$) for all three objects. For \gfiftyfour, even at $0.2 \pc$ from the heating source the silicate dust mass is $0.09 \msun$, although due to the complex geometry of this object we are unable to seriously constrain the mass with our models.}

In contrast to the grain sizes inferred from observations, CCSNe dust formation models generally do not predict significant masses of dust grains with sizes $> 0.1 \um$. \citet{nozawa2003} found that the combined size distribution of all species approximately followed a $-3.5$ power law above $0.1 \um$, which would imply the majority of the mass is contained in these sizes, but more recent studies (e.g. \citealt{bianchi2007,sarangi2015,biscaro2016}) have not necessarily reproduced this. \citet{bocchio2016} did produce significant $> 0.1 \um$ grain populations in their dust formation models, but not in the case of the Crab Nebula, whereas we find large grains are necessary to reproduce the Crab IR SED. { The grain size distributions presented in \citet{marassi2019} do not extend beyond a few $\times 0.1 \um$, and then only for carbon grains. \citet{omand2019} found that, for the range of pulsar parameters they investigated, the presence of a PWN reduces the grain size of newly formed dust from $\sim 0.01 \um$ to $\sim 10^{-3} \um$, whereas the PWNe we investigate have typical grain sizes well above even their non-PWN model.} Dust formation models also generally find that the size distribution is log-normal - while, due to the uncertainties, we are unable to rule this scenario out, our results seem to favour power law distributions, particularly in the case of \geleven \, and \gfiftyfour \, where both small and large grains must contribute to the SED. An initially log-normal size distribution can be converted to a power law by further grain processing, in particular grain-grain collisions \citep{jones1996}.

Grain size is a critical parameter in determining the destruction of dust by sputtering. Studies of dust survival rates in SNRs after processing by the reverse shock have often taken the initial size distribution from the dust formation models previously discussed \citep{nozawa2007,biscaro2016,bocchio2016} - if these models are underestimating the relative importance of large grains, the derived destruction rates will be overestimated. \citet{nozawa2007} found that grains of size $\gtrsim 0.2 \um$ were essentially unaffected by sputtering and survived intact into the ISM. From our results, this would imply that at least $\sim 50 \%$, and possibly up to essentially $100 \%$, of the dust mass can survive, compared to literature values ranging from $\sim 10\%$ \citep{micelotta2016,biscaro2016} down to $<1\%$ \citep{bocchio2016}. The grains would also be more resistant to subsequent destruction in the ISM. However, this assumes the dust is destroyed only by sputtering - for large grains, grain-grain collisions may be a significant additional destruction mechanism \citep{kirchschlager2019}.

To summarise, we have used physical dust heating models for five Galactic PWNe to fit the observed dust SEDs with multiple single-grain size emission components, determining both the dust mass and the grain size distribution. Our dust masses { generally} agree with previous studies of the same objects using different methodologies, confirming that { CCSNe are potentially significant producers of newly-formed dust}. In all cases, we find that grains with radii $\ge 0.1 \um$ make up the vast majority of the total dust mass, with strong evidence for the presence of micron-sized grains, which have previously been proposed to exist in SN2010jl \citep{gall2014}, the Crab Nebula \citep{owen2015}, Cas A \citep{bevan2016} and SN1987A \citep{wesson2015}. With the addition of the four other PWNe from this paper, every SNR for which the grain size has been investigated seems to contain large grains, with important consequences for the injection into the ISM, and the subsequent survival, of the newly-formed ejecta dust.

\section*{Acknowledgements}

{ We are grateful to the referee for suggestions which significantly improved this paper.} FDP is supported by the Science and Technology Facilities Council. MJB acknowledges support from the European Research Council (ERC) grant SNDUST ERC-2015-AdG-694520. IDL gratefully acknowledges the support of the Research Foundation -- Flanders (FWO). HC acknowledges support from the ERC grant COSMICDUST ERC-2014-CoG-647939.




\bibliographystyle{mnras}
\bibliography{crabdust}




\bsp	
\label{lastpage}
\end{document}